\documentclass[twocolumn,resetfootnote,trackchanges]{aastex701}

\shorttitle{NPE for Tomographic Mass Mapping}
\shortauthors{White et al.}

\usepackage{graphicx}
\usepackage{amsmath}
\usepackage{anyfontsize}
\usepackage{subcaption}
\usepackage{tikz, pgfplots}
\usetikzlibrary{positioning,shapes.geometric,bending,fit}
\pgfplotsset{compat=1.18}
\usepackage[nameinlink,capitalize,noabbrev]{cleveref}
\creflabelformat{equation}{#2#1#3}
\crefname{appendix}{Appendix}{Appendices}
\Crefname{appendix}{Appendix}{Appendices}
\usepackage{caption}
\usepackage{etoolbox}
\makeatletter
\apptocmd\@sect{%
  \cref@constructprefix{#1}{\cref@result}%
  \@ifundefined{cref@#1@alias}%
    {\def\@tempa{#1}}%
    {\def\@tempa{\csname cref@#1@alias\endcsname}}%
  \protected@edef\cref@currentlabel{%
    [\@tempa][\arabic{#1}][\cref@result]%
    \csname p@#1\endcsname\csname the#1\endcsname}%
}{}{\fail}
\makeatother

\usepackage{todonotes}
\setuptodonotes{inline}

\usepackage{multirow}
\usepackage{booktabs}
\usepackage{nicefrac}

\newcommand{\padvert}{\thinspace \vert \thinspace}
\definecolor{browngray}{RGB}{143, 108, 39}
\definecolor{deepgreen}{RGB}{28, 105, 43}
\definecolor{mediumteal}{RGB}{9, 123, 125}
\definecolor{pinkgray}{RGB}{122, 104, 119}

\begin{document}

\title{Neural Posterior Estimation for Tomographic Weak Lensing Mass Mapping}

\author[orcid=0000-0001-5535-0452]{Tim White}
\affiliation{Department of Statistics, University of Michigan}
\email{twhit@umich.edu}

\author[orcid=0009-0005-6557-582X]{Shreyas Chandrashekaran}
\affiliation{Department of Computer Science and Engineering, University of Michigan}
\email{shreyasc@umich.edu}

\author[orcid=0000-0001-8868-0810]{Camille Avestruz}
\affiliation{Department of Physics, University of Michigan}
\email{cavestru@umich.edu}

\author[orcid=0000-0002-1472-5235]{Jeffrey Regier}
\affiliation{Department of Statistics, University of Michigan}
\email{regier@umich.edu}

\collaboration{all}{The LSST Dark Energy Science Collaboration}

\begin{abstract}
Weak gravitational lensing shear and convergence trace the distribution of baryonic and dark matter across space, making them a powerful probe of cosmic structure. Inferring shear and convergence from images is a challenging inverse problem. The prevailing approach to this task estimates shear from weighted averages of galaxy ellipticities, calibrates these estimates to account for systematic biases, and transforms them to reconstruct convergence, a multistage procedure that requires substantial computational resources and meticulous handling of statistical uncertainties. As an alternative, we propose a probabilistic approach to field-level weak lensing inference in which we train a deep neural network to directly map a multiband image to a variational distribution over the underlying tomographic shear and convergence fields. This neural posterior estimation (NPE) procedure implicitly marginalizes over nuisance variables in the cosmological forward model and does not require evaluating the likelihood function. It is also amortized, so it enables rapid posterior inference for astronomical surveys once the neural network is trained. When evaluated on synthetic images from the LSST-DESC DC2 Simulated Sky Survey, NPE produces well-calibrated variational distributions for shear and convergence that are consistent with the ground truth. We describe how maps sampled from these variational distributions could be used in a subsequent simulation-based inference procedure to approximate the posterior distribution over cosmological parameters.
\end{abstract}


\section{Introduction} \label{sec:introduction}

Light from distant galaxies is deflected by matter along the telescope's line of sight, which modifies their observed shapes and brightnesses in astronomical images. This phenomenon is known as weak gravitational lensing. Weak lensing is characterized by shear, which describes anisotropic distortion, and convergence, which describes isotropic magnification. These quantities trace the projected distribution of matter across space, making weak lensing a powerful probe of the Universe's large-scale structure \citep{bartelmann2001weak, kilbinger2015cosmology, dodelson2017gravitational, prat2025weak}.

The next generation of astronomical surveys, including the Rubin Observatory's Legacy Survey of Space and Time (LSST) \citep{ivezic2019lsst}, the Euclid Space Telescope \citep{laureijs2011euclid}, and the Nancy Grace Roman Space Telescope \citep{dore2019wfirst}, will image billions of galaxies across thousands of square degrees of the night sky. These Stage IV surveys offer an opportunity to constrain cosmological parameters with unprecedented precision based on measurements of weak gravitational lensing \citep{mandelbaum2018lsst}. To take full advantage of the opportunity, shear and convergence estimation algorithms must efficiently process petabytes of image data, limit information loss, and quantify the uncertainty of their estimates.

Current approaches to weak lensing inference follow a multistage pipeline. First, images are processed to perform object detection, deblending, and measurement. Next, shear is estimated from galaxy ellipticities, typically using self-calibrating methods that estimate the shear response through artificial shearing (Metadetection; \cite{sheldon2023metadetection}) or analytical modeling (AnaCal; \cite{li2024differentiable}). Convergence is then reconstructed from shear estimates, either through direct inversion methods \citep{kaiser1993mapping} or probabilistic reconstruction techniques \citep{schneider2017probabilistic, jeffrey2021dark, remy2023probabilistic, whitney2024generative, leterme2026plug}. Finally, cosmological parameters are inferred from summary statistics of the shear and convergence fields, such as two-point correlation functions \citep{abbott2022dark, prat2022catalog} or higher-order statistics \citep{gatti2022dark, jeffrey2024dark, makinen2025hybrid}.

Each stage of this pipeline requires careful propagation of statistical uncertainties, and information about the underlying shear and convergence fields is typically lost when images are compressed, first to galaxy catalogs and then to summary statistics. Recent work has shown that field-level inference, which uses the full convergence maps rather than summary statistics, can yield tighter constraints on cosmological parameters \citep{sharma2024comparative, lanzieri2024optimal, boruah2024map, fiedorowicz2022karmma, zhou2024accurate}.

Several works have explored Bayesian inference of shear and convergence from images \citep{miller2007bayesian, bernstein2014bayesian, schneider2015hierarchical, bernstein2016accurate, heavens2018bayesian, congedo2024euclid, sallaberry2025scalable}. Many of these works posit a forward model of lensed galaxy fields and use Markov chain Monte Carlo to generate posterior samples under this model. This approach requires repeated evaluation of a likelihood function over all latent variables in the forward model, including the intrinsic properties of the imaged galaxies and the parameters of spatially varying backgrounds and point spread functions. While these methods provide principled uncertainty quantification, the computational cost of repeated likelihood evaluation makes their application to modern surveys challenging.

As an alternative, we propose to infer shear and convergence maps from multiband coadded images using neural posterior estimation (NPE), a type of simulation-based inference. This procedure involves training a neural network to map an image to a variational distribution over lensing maps for several redshift bins. NPE is amortized — once the neural network has been trained, it can estimate shear and convergence in a single forward pass.

Our approach differs from existing methods in several respects. The network maps images directly to the underlying lensing fields without first compressing the image pixels into galaxy catalogs. Additionally, NPE is likelihood-free, unlike MCMC-based methods. Finally, NPE targets the posterior distribution over tomographic lensing maps, unlike neural-network-based point estimators of shear \citep{tewes2019weak, ribli2019galaxy, zhang2024forklens}.

In this work, we investigate the feasibility of NPE as a field-level weak lensing inference method by applying it to the LSST Dark Energy Science Collaboration's DC2 Simulated Sky Survey \citep{abolfathi2021desc, abolfathi2021lsst}. The DC2 lensing maps, object catalogs, and coadded images can be viewed as samples from an implicit probabilistic model (\cref{sec:dc2}). We formulate the NPE objective and describe our choice of variational family and neural network architecture (\cref{sec:npe}). We train the network on images covering a contiguous DC2 subregion (\cref{sec:traineval}) and evaluate its accuracy and calibration on a different subregion (\cref{sec:results}).

In NPE, the network is trained exclusively on simulated images; we discuss strategies for mitigating potential simulator misspecification, which would enable the trained network to be applied to real astronomical surveys (\cref{sec:discussion}). Additionally, we outline how conditional flow matching could be used to fit more flexible posterior approximations. Finally, we describe how mass maps sampled from our fitted variational distributions could be passed to a subsequent neural simulation-based inference procedure that approximates the posterior distribution over $\Lambda$CDM model parameters. When connected, the two procedures would constitute an end-to-end probabilistic pipeline from images to cosmological parameters.

\section{The LSST DESC DC2 Simulated Sky Survey} \label{sec:dc2}

The Vera C. Rubin Observatory Legacy Survey of Space and Time (LSST) commenced in early 2026 \citep{ivezic2019lsst}, with the first data release expected in 2028 \citep{guy2024rubin}. It will capture high-resolution images of billions of galaxies across 18,000 square degrees of the sky. To ensure that software pipelines are prepared to process and analyze these images, the LSST Dark Energy Science Collaboration created a synthetic sky survey as part of its second data challenge (DC2). \cite{korytov2019cosmodc2}, \cite{abolfathi2021desc}, and \cite{abolfathi2021lsst} describe the DC2 Simulated Sky Survey in detail, and \cite{duan2026neural} concisely summarize its major components. Below, we describe the DC2 data products that are relevant to weak lensing inference (\cref{subsec:dc2mapsandimages}) and cast them as a realization of an implicit probabilistic model (\cref{subsec:dc2notation}).

\subsection{Coadded images and lensing maps} \label{subsec:dc2mapsandimages}

The galaxies imaged in DC2 are from cosmoDC2, a synthetic extragalactic catalog \citep{korytov2019cosmodc2}. To construct cosmoDC2, a base catalog of unlensed galaxies is formed from the outputs of the Outer Rim N-body simulation \citep{heitmann2019outer} and the UniverseMachine catalog \citep{behroozi2019universemachine}. Next, the galaxies are lensed based on shear and convergence maps. These maps are created by projecting particles onto discrete redshift shells from the Outer Rim lightcone, mapping the surface density of each shell, and tracing photon paths through the shells. The shear and convergence values applied to each galaxy are determined by interpolating these maps to the galaxy's redshift and sky coordinates. Finally, the catalog is augmented with additional properties after matching the lensed galaxies to those from the Galacticus library \citep{benson2012galacticus}. \cref{fig:dc2galaxyproperties} displays the distributions of several properties for a random sample of cosmoDC2 galaxies.

Given the cosmoDC2 catalog (along with separate catalogs of Milky Way stars and Type Ia supernovae), the imSim software package simulates raw $4000{\times}4000$-pixel images in six optical bands ($ugrizy$), which together cover 300 square degrees of the sky. Objects are rendered in the images using the GalSim package \citep{rowe2015galsim}. The imSim package mimics both the LSST observing cadence and the sensor and readout effects of the LSST Camera \citep{abolfathi2021desc}.

The LSST Data Release Production (DRP) pipeline processes the raw images by performing tasks such as instrument signature removal, background estimation, and source detection and deblending. Once the images are processed and characterized, the DRP pipeline calibrates them against a reference catalog to produce calibrated exposures. Then, coadded images (i.e., ``coadds'') are generated by resampling the calibrated exposures on a common grid defined by tracts and patches. Each tract comprises $7{\times}7$ patches, and each patch comprises $4100{\times}4100$ pixels. The pixel scale is 0.2 arcseconds.

Two tracts of coadds are generated for DC2. These coadds cover approximately five square degrees of the sky, a small fraction of the coverage of the raw images. Despite this, we analyze the coadds instead of the raw images, as is standard in weak lensing analyses \citep{armstrong2024little}. In doing so, we leverage the coadds' high signal-to-noise ratio and avoid the need to combine inferences from overlapping raw images.

\begin{figure*}
  \centering
  \begin{subfigure}[b]{0.26\textwidth}
    \vspace*{\fill}
    \includegraphics[width=\textwidth]{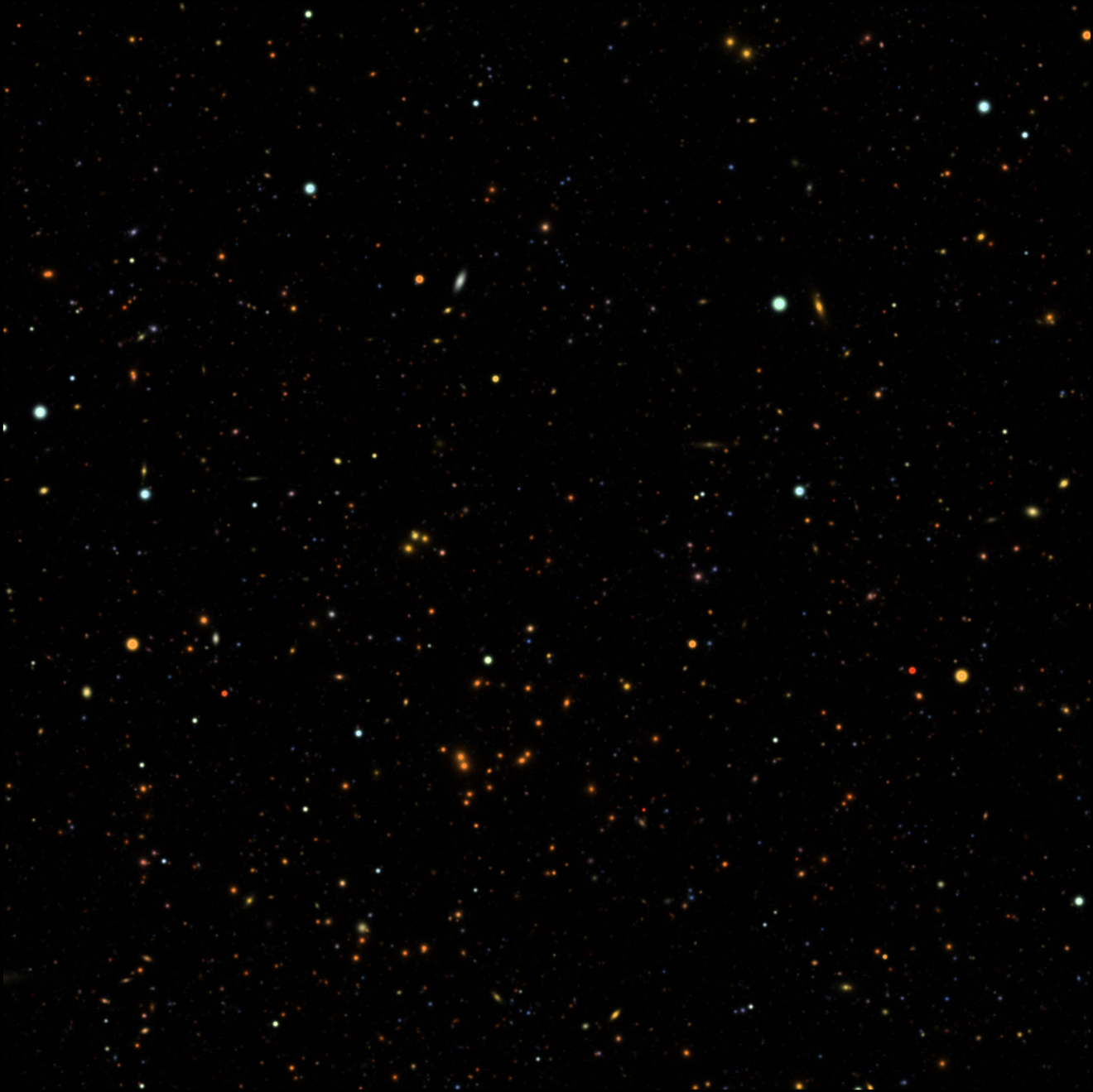}
    \vspace*{\fill}
  \end{subfigure}
  \hfill
  \begin{subfigure}[b]{0.35\textwidth}
    \vspace*{\fill}
    \includegraphics[width=\textwidth]{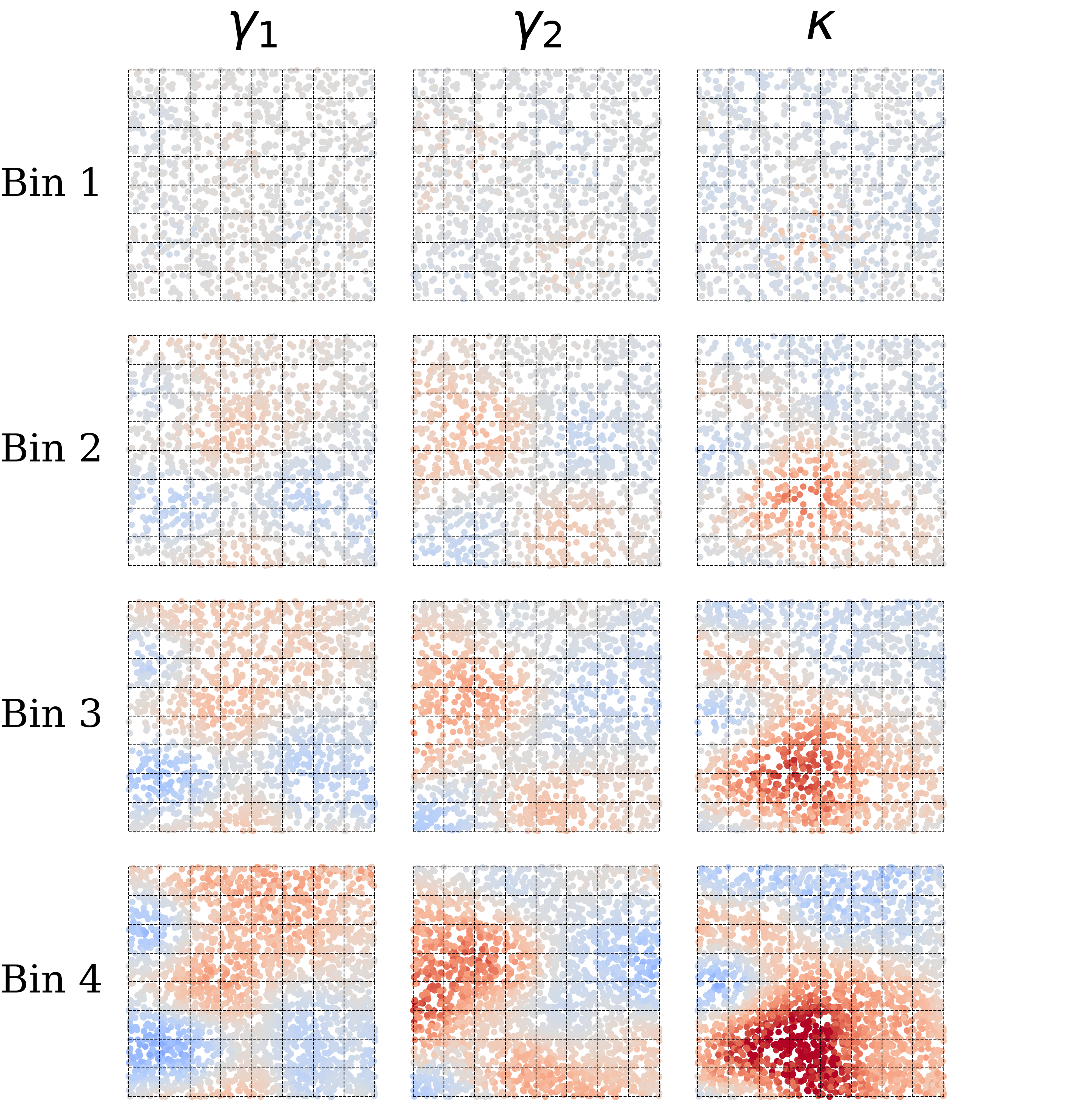}
  \end{subfigure}
  \hfill\hspace{-45pt}
  \begin{subfigure}[b]{0.35\textwidth}
    \vspace*{\fill}
    \includegraphics[width=\textwidth]{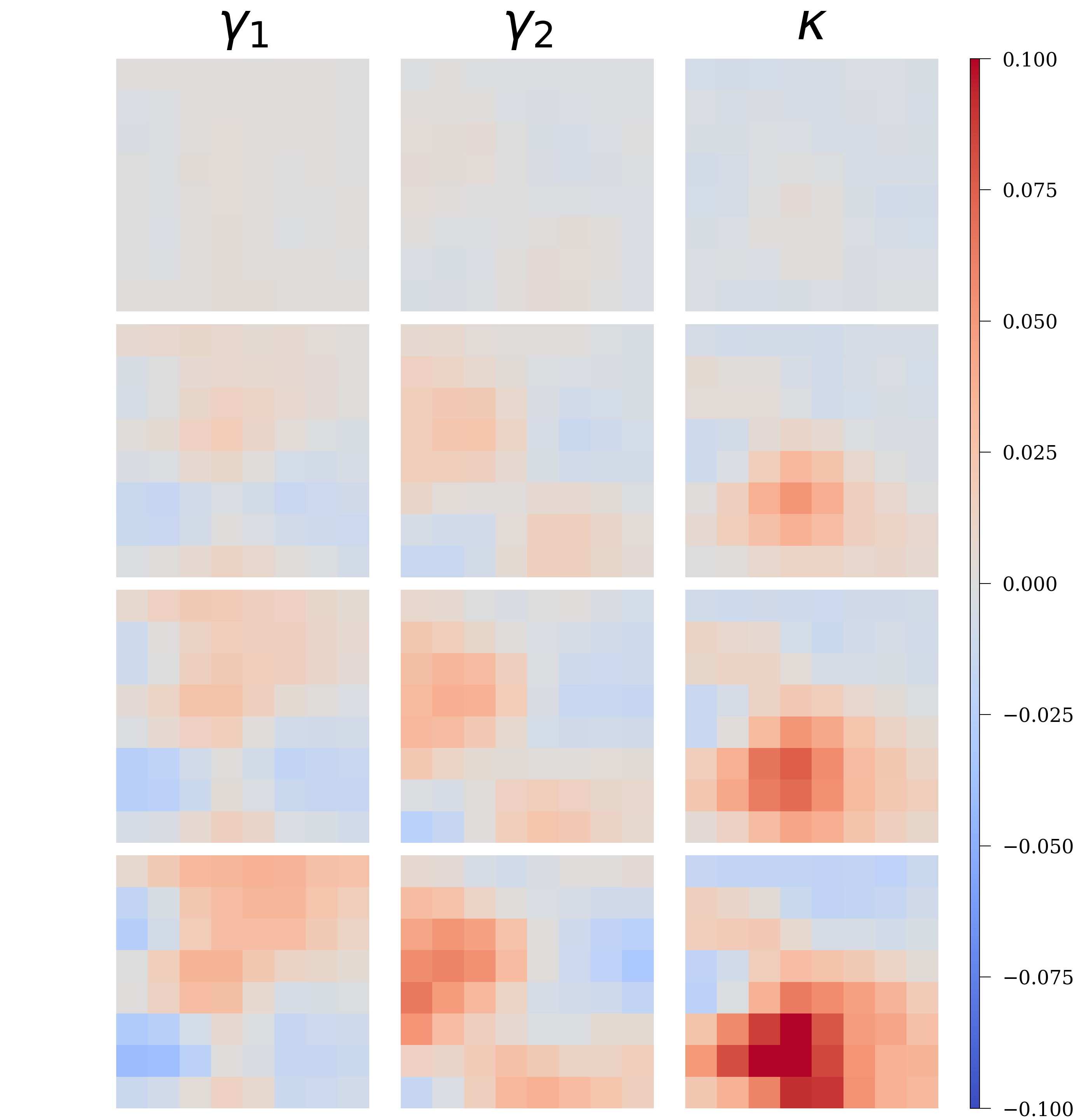}
  \end{subfigure}
  \caption{Left: A $2048{\times}2048$-pixel DC2 image formed using the $g$, $r$, and $i$ optical bands. Center: Shear and convergence of cosmoDC2 galaxies whose centroids lie within the image boundary. Each galaxy is assigned to one of four redshift bins defined by the intervals (0.007, 0.763), (0.763, 1.120), (1.120, 1.593), (1.593, 3.123). Right: Tomographic shear ($\gamma_1, \gamma_2$) and convergence ($\kappa$) maps constructed by averaging each quantity over $256{\times}256$-pixel tiles within each redshift bin.}
  \label{fig:dc2imagegroundtruth}
\end{figure*}

A single patch-level coadd is too large to store in memory during network training, so we partition each one into four $6{\times}2048{\times}2048$-pixel coadds. There are 392 of these in total; we display one in the left panel of \cref{fig:dc2imagegroundtruth}. All mentions of coadds or images in the remainder of this paper refer to these $6{\times}2048{\times}2048$-pixel objects, unless otherwise specified.

While these images are the inputs to the neural network in our neural posterior estimation procedure, training the network also requires the Outer Rim lensing maps described above. To the best of our knowledge, these maps were not published. However, it is possible to approximate them using the per-galaxy shear and convergence values reported in cosmoDC2.

\cref{fig:dc2imagegroundtruth} demonstrates this approximation procedure for a single image. The center panel displays the shear and convergence at the positions of a random sample of galaxies whose centroids lie within the image boundary. Each galaxy is assigned to one of $B{=}4$ redshift bins defined by the intervals (0.007, 0.763), (0.763, 1.120), (1.120, 1.593), (1.593, 3.123). The bin thresholds are the minimum, first quartile, median, third quartile, and maximum redshift among galaxies detected by the LSST Science Pipelines \citep{abolfathi2021desc}.

For each redshift bin, we compute the mean shear and convergence in $S{\times}S$-pixel ``tiles'' defined on the same coordinate space as the image. The tile side length $S$ must be a divisor of the image dimension (2048, in our case). It should be chosen in tandem with the redshift bin count $B$ such that (i) shear and convergence are each approximately uniform within each tile and (ii) the galaxy count per tile is large enough to ensure that sampling noise in the shear and convergence averages is small.

This spatial binning procedure produces tomographic shear and convergence maps of size $B{\times}H{\times}W$, where $H{=}W{=}2048{/}S$. In this work, we use $B{=}4$ redshift bins and a tile side length of $S{=}256$, so $H{=}W{=}8$. The right panel of \cref{fig:dc2imagegroundtruth} displays these maps. We denote the tomographic convergence map for a single image as
\begin{equation}
    \kappa = \{\kappa^{(b,t)}: b{=}1,...,B; \thinspace t{=}1,...,HW\},
\end{equation}
where $b$ and $t$ index the bins and tiles, respectively. We use the same notation for the shear maps $\gamma_1$ and $\gamma_2$.

\subsection{An implicit probabilistic model} \label{subsec:dc2notation}

The lensing maps, stellar and galactic catalogs, and coadds described in \cref{subsec:dc2mapsandimages} can be viewed as a single sample from a probabilistic generative model specified implicitly by the DC2 simulator. Let $z := \{\gamma_1, \gamma_2, \kappa\}$ denote the tomographic shear and convergence maps underlying a single image. Thus, $z$ has dimension $3{\times}B{\times}H{\times}W$. Given the fiducial DC2 cosmology and the hyperparameters of the Outer Rim simulation, these maps are generated stochastically according to an implicit prior $p(z)$.

Let $z_{\text{nuisance}}$ denote all other latent variables encoded in the image, including the properties of the imaged light sources and the characteristics of the spatially varying background, point spread function (PSF), and pixel-level noise. The latter characteristics could alternatively be viewed as fixed inputs to the forward model. The shear and convergence maps $z$ do not affect the background, PSF, or image noise, but they do alter the intrinsic shape and flux of each galaxy whose centroid falls within the image boundary. Thus, the latent variables $z_{\text{nuisance}}$ can be viewed as samples from an implicit prior $p(z_{\text{nuisance}} \padvert z)$ that reflects this dependence. Since each image depicts thousands of light sources, the dimension of $z_{\text{nuisance}}$ is on the order of thousands, if not tens of thousands.

Let $x$ denote the $6{\times}2048{\times}2048$ image generated from $z$ and $z_{\text{nuisance}}$ according to the forward model $p(x \padvert z, z_{\text{nuisance}})$. This forward model encompasses both the simulation of raw exposures using the GalSim and imSim packages and the calibration and coaddition of these raw exposures by the LSST Data Release Production pipeline.

\begin{figure*}
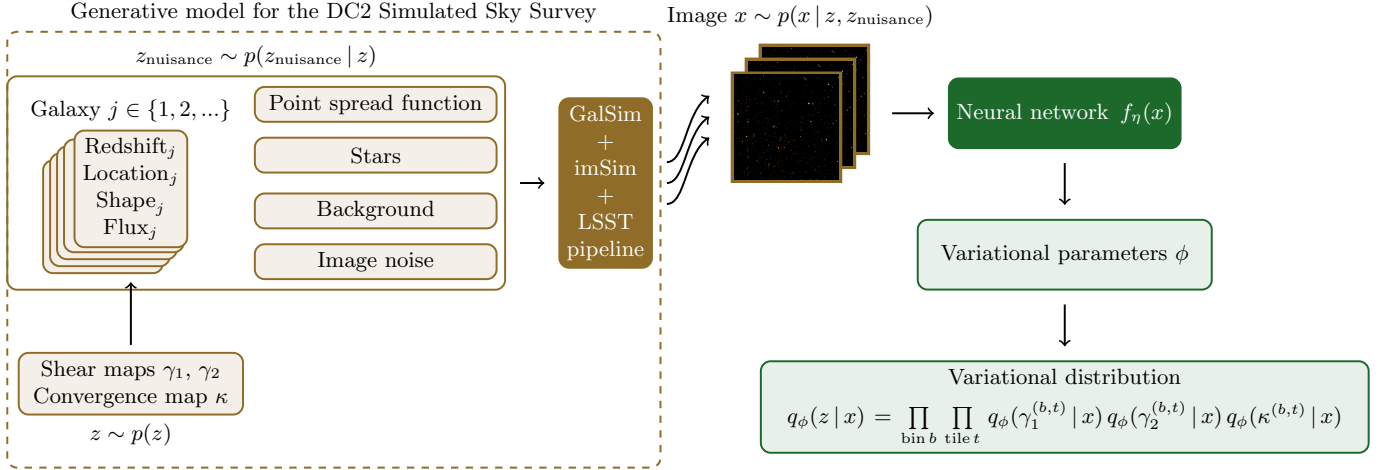

    \resizebox{\textwidth}{!}{%
    \begin{tikzpicture}[
    dc2/.style={
    rectangle, thick, dashed, rounded corners, minimum width=6cm, minimum height=6.25cm, draw=browngray
    },
    outlineinput/.style={
    rectangle, thick, rounded corners, minimum width=2.5cm, minimum height=2.25cm, draw=browngray
    },
    inputbox/.style={
    rectangle, thick, rounded corners, minimum width=1.5cm, minimum height=0.5cm, text centered, draw=browngray, fill=browngray!10
    },
    galsim/.style={
    rectangle, thick, rounded corners, minimum width=1cm, minimum height=1cm, text centered, draw=browngray, fill=browngray!10
    },
    image/.style={
    rectangle, thick, line width=0.1cm, draw=browngray
    },
    inferencenetwork/.style={
    rectangle, thick, rounded corners, minimum width=1cm, minimum height=1cm, text centered, draw=deepgreen, fill=deepgreen!10
    },
    outputbox/.style={
    rectangle, thick, rounded corners, minimum width=1.5cm, minimum height=1cm, text centered, draw=deepgreen, fill=deepgreen!10
    },
    ]
    
    \node (galaxies) {
    \begin{tikzpicture}
        \node[inputbox, text height=1.4cm, text width=1.5cm] at (-0.2, -0.2) {};
        \node[inputbox, text height=1.4cm, text width=1.5cm] at (-0.1, -0.1) {};
        \node[inputbox, text height=1.4cm, text width=1.5cm] at (0, 0) {};
        \node[inputbox, text height=1.4cm, text width=1.5cm] at (0.1, 0.1) {};
        \node[inputbox, text width=1.4cm, label=above:{Galaxy $j\in\{1,2,...\}$}] (galaxy) at (0.2,0.2) {$\text{Redshift}_j$\\
        $\text{Location}_j$\\
        $\text{Shape}_j$\\
        $\text{Flux}_j$};
    \end{tikzpicture}
    };
    \node[inputbox, label=below:{$z \sim p(z)$}, text width=3cm] (shearconv) [below=of galaxies] {Shear maps $\gamma_1$, $\gamma_2$\\
    Convergence map $\kappa$};
    \node[inputbox, text width=3.25cm] (stars) [right=of galaxies, xshift=-0.9cm,  yshift=0.4cm] {Stars};
    \node[inputbox, text width=3.25cm] (psf) [above=0.2cm of stars] {Point spread function};
    \node[inputbox, text width=3.25cm] (background) [right=of galaxies, xshift=-0.9cm, yshift=-0.4cm] {Background};
    \node[inputbox, text width=3.25cm] (noise) [below=0.2cm of background] {Image noise};
    \node[outlineinput, label=above:{$z_{\text{nuisance}} \sim p(z_{\text{nuisance}} \padvert z)$}, fit={(galaxies) (background) (psf) (noise)}] (nuisance) {};
    
    \node[galsim, fill=browngray, text=white, align=center] (galsim) [right=of nuisance, xshift=-0.25cm] {GalSim\\
    +\\
    imSim\\
    +\\
    LSST\\
    pipeline};

    \node[dc2, fit={(galaxies) (background) (psf) (galsim) (shearconv)}, label=above:{Generative model for the DC2 Simulated Sky Survey}] (dc2) {};
    
    \node (image) [right=1cm of galsim, yshift=1cm, label=above:{Image $x \sim p(x \padvert z, z_{\text{nuisance}})$}] {
    \begin{tikzpicture}
        \node[image, inner sep=0pt] at (0.2, 0.2) {\includegraphics[width=1.5cm]{figures/dc2image.png}};
        \node[image, inner sep=0pt] at (0, 0) {\includegraphics[width=1.5cm]{figures/dc2image.png}};
        \node[image, inner sep=0pt] at (-0.2, -0.2) {\includegraphics[width=1.5cm]{figures/dc2image.png}};
    \end{tikzpicture}
    };
    \node[inferencenetwork, fill=deepgreen, text=white] (inferencenetwork) [right=1cm of image] {Neural network \thinspace $f_\eta(x)$};
    
    \node[outputbox] (variationalparams) [below=of inferencenetwork, text centered, text width=4cm] {Variational parameters $\phi$};
    \node[outputbox] (posterior) [below=of variationalparams, text centered, text width=8.5cm] {Variational distribution \\[5pt]{$q_\phi(z \padvert x) = \underset{\text{bin} \thinspace b}{\prod} \thinspace\thinspace \underset{\text{tile} \thinspace t}{\prod} \thinspace\thinspace q_{\phi}(\gamma_1^{(b,t)} \padvert x) \thinspace q_{\phi}(\gamma_2^{(b,t)} \padvert x) \thinspace q_{\phi}(\kappa^{(b,t)} \padvert x)$}};
    
    \draw[->, thick, shorten <=0.1cm] (shearconv.north) to (galaxies.south);
    \draw[->, thick, shorten <=0.2cm, shorten >=0.2cm] (nuisance.east) to (galsim.west);
    \draw[->, thick, shorten <=0.2cm, shorten >=0.2cm, out=0, in=180] ([yshift=-0.3cm]galsim.east) to ([yshift=-0.3cm]image.west);
    \draw[->, thick, shorten <=0.2cm, shorten >=0.2cm, out=0, in=180] ([yshift=0cm]galsim.east) to ([yshift=0cm]image.west);
    \draw[->, thick, shorten <=0.2cm, shorten >=0.2cm, out=0, in=180] ([yshift=0.3cm]galsim.east) to ([yshift=0.3cm]image.west);
    \draw[->, thick, shorten <=0.2cm, shorten >=0.2cm] (image.east) to (inferencenetwork.west);
    \draw[->, thick, shorten <=0.2cm, shorten >=0.2cm] (inferencenetwork.south) to (variationalparams.north);
    \draw[->, thick, shorten <=0.2cm, shorten >=0.2cm] (variationalparams.south) to (posterior.north);
    \end{tikzpicture}}
    \caption{Our inference framework. Lensing maps $z = \{\gamma_1, \gamma_2, \kappa\}$, non-lensing variables $z_{\text{nuisance}}$, and images $x$ are generated by the DC2 simulator (\cref{subsec:dc2mapsandimages}), which can be viewed as an implicit generative model (\cref{subsec:dc2notation}). Given the images, a neural network $f_\eta$ approximates the posterior distribution over lensing maps via neural posterior estimation (\cref{subsec:npeobjective}). The network is trained to map each image to the parameters $\phi$ of a distribution $q_\phi(z \padvert x)$ belonging to a spatially factorized variational family (\cref{subsec:variationalfamily}). Our network $f_\eta$ is a deep residual network with weights $\eta$ (\cref{subsec:trainingarchitecture}).}
    \label{fig:inferencediagram}
\end{figure*}

The left panel of \cref{fig:inferencediagram} illustrates how the joint distribution $p(z, z_{\text{nuisance}}, x)$ factorizes as $p(z) p(z_{\text{nuisance}} \padvert z) p(x \padvert z, z_{\text{nuisance}})$. Any latent quantities excluded from this panel, such as the intrinsic shape and flux of each galaxy, are implicitly absorbed by $z_{\text{nuisance}}$. The DC2 lensing maps, catalogs, and coadds are generated through repeated calls to this generative model, as illustrated by the replicated arrows in \cref{fig:inferencediagram}.

\section{Neural posterior estimation} \label{sec:npe}

Given an image $x$, we aim to infer the posterior distribution $p(z \padvert x)$ over tomographic shear and convergence maps. This distribution, which is a marginal of the joint posterior $p(z, z_{\text{nuisance}} \padvert x)$, can be expressed as
\begin{align}
\begin{split}
    p(z \padvert x) &= \int p(z, z_{\text{nuisance}} \padvert x) dz_{\text{nuisance}} \\
    &= \int \frac{p(z) p(z_{\text{nuisance}} \padvert z) p(x \padvert z, z_{\text{nuisance}})}{p(x)} dz_{\text{nuisance}}.
\end{split}
\end{align}
One approach to inferring $p(z \padvert x)$ is to sample from $p(z, z_{\text{nuisance}} \padvert x)$ using Markov chain Monte Carlo and discard the sampled nuisance variables. However, this procedure requires the repeated evaluation of the likelihood $p(x \padvert z, z_{\text{nuisance}})$, which is computationally demanding since $x$, $z$, and $z_{\text{nuisance}}$ are all high-dimensional. Also, for highly realistic simulations like DC2, writing $p(x \padvert z, z_{\text{nuisance}})$ in closed form may not be feasible.

\subsection{An amortized, likelihood-free objective} \label{subsec:npeobjective}

To circumvent the computational challenges described above, we approximate $p(z \padvert x)$ with a variational distribution $q_\phi(z \padvert x)$ parametrized by a vector $\phi \in \Phi$. We discuss our choice of $\Phi$, which defines a family of variational distributions, in \cref{subsec:variationalfamily}.

We fit $q$ using neural posterior estimation (NPE), a type of amortized variational inference. In NPE, instead of optimizing the variational parameters $\phi$ separately for each image $x$, we train a neural network to map any $x$ to a corresponding $\phi$ — i.e., $\phi = f_\eta(x)$, where the function $f_\eta$ is a neural network with weights $\eta$. Thus, the computational cost of solving this inference problem is amortized across $(z,x)$ pairs sampled from the DC2 generative model. The weights $\eta$ are selected to minimize
\begin{equation} \label{eq:npeobjective}
    \mathcal{L}(\eta) = E_{p(x)} \left[D_{\text{KL}}\left(p(z \padvert x) \thinspace\Vert\thinspace q_{f_\eta(x)} (z \padvert x) \right)\right],
\end{equation}
the expected forward Kullback-Leibler divergence of the true marginal posterior from the variational distribution. Substituting the definition of the KL divergence into \cref{eq:npeobjective} yields the simplified objective
\begin{equation} \label{eq:npeobjective_simple}
    \mathcal{L}(\eta) = - E_{p(z,x)} \left[\log q_{f_\eta(x)}(z \padvert x)\right] + C,
\end{equation}
where $C$ is a constant with respect to $\eta$. The expectation is taken with respect to $(z,x)$ pairs obtained by sampling from $p(z, z_{\text{nuisance}}, x)$ and discarding the non-lensing variables $z_{\text{nuisance}}$.

We use stochastic gradient descent to obtain the neural network weights $\eta$ that minimize \cref{eq:npeobjective_simple}. This involves computing Monte Carlo approximations of the gradient
\begin{equation}
    \nabla \mathcal{L}(\eta) = - E_{p(z,x)} \left[\nabla \log q_{f_\eta(x)}(z \padvert x)\right]
\end{equation}
using batches of $(z,x)$ pairs sampled from $p(z,x)$. The weights $\eta$ are iteratively updated based on these approximate gradients. We use a batch size of one, as we found that at most one image can be held in memory along with the network weights.

The NPE loss function is convenient for three reasons. First, it is likelihood-free, as it does not require evaluating $p(x \padvert z, z_{\text{nuisance}})$. Second, while it requires sampling $z_{\text{nuisance}}$ from the generative model, it otherwise marginalizes over these nuisance variables, as they do not appear in the integrand. Finally, it enables statistically consistent estimation of $p(z \padvert x)$, provided that the variational family is sufficiently flexible \citep{frazier2024statistical}.

\subsection{A spatially factorized variational family} \label{subsec:variationalfamily}

In this work, we evaluate the performance of NPE for a family of variational distributions that factorize over redshift bins, tiles, and the three lensing variables. Variational families with this structure are often called mean-field families \citep{zhang2019advances}. The results presented in \cref{sec:results} establish a baseline for future work that will posit a more flexible variational family.

We define a family $\mathcal{Q} := \{q_\phi(z \padvert x) : \phi \in \Phi\}$ comprising variational distributions of the form
\begin{equation} \label{eq:variationalfamily}
    q_\phi(z \padvert x) = \prod_{b=1}^B \prod_{t=1}^{HW} q_\phi(\gamma_1^{(b,t)} \padvert x) q_\phi(\gamma_2^{(b,t)} \padvert x) q_\phi(\kappa^{(b,t)} \padvert x),
\end{equation}
where the products are taken over $B$ redshift bins and $H{\times}W$ tiles, respectively. For each bin $b$ and tile $t$, we model $\gamma_1^{(b,t)}$, $\gamma_2^{(b,t)}$, and $\kappa^{(b,t)}$ as independent Gaussian random variables. While in \cref{eq:variationalfamily} we index each of these variational factors by $\phi$ to avoid notational clutter, each factor is actually parametrized by a subset of $\phi$ comprising one mean parameter and one variance parameter. The entire vector $\phi$ is formed by flattening an array of size $B {\times} H {\times} W {\times} 3 {\times} 2$, whose final two dimensions represent the number of lensing variables and the number of Gaussian distribution parameters per lensing variable, respectively.

While this mean-field variational family models $\gamma_1^{(b,t)}$, $\gamma_2^{(b,t)}$, and $\kappa^{(b,t)}$ as independent (conditional on the image), we expect these variables to exhibit non-zero auto- and cross-correlation in the exact posterior. Later, we will propose a more flexible class of variational distributions (\cref{subsec:flowmatching}) from which spatially dependent shear and convergence maps can be sampled, summarized, and used to constrain cosmological parameters (\cref{subsec:cosmologicalparams}).

Although its dependence structure does not match that of the exact posterior, our variational family nevertheless provides a valuable approximation of it. For one thing, maps sampled from the fitted variational distributions can still exhibit non-zero auto- or cross-correlation at large scales, even though shear and convergence are modeled as independent at the tile level. The neural network's receptive field spans many tiles, so the per-tile means it predicts can carry spatial structure inherited from the underlying maps. Additionally, the consistent marginalization property of the NPE objective guarantees that the variational factors $q_\phi(\gamma_1^{(b,t)} \padvert x)$, $q_\phi(\gamma_2^{(b,t)} \padvert x)$, and $q_\phi(\kappa^{(b,t)} \padvert x)$ that minimize \cref{eq:npeobjective_simple} coincide exactly with $p(\gamma_1^{(b,t)} \padvert x)$, $p(\gamma_2^{(b,t)} \padvert x)$, and $p(\kappa^{(b,t)} \padvert x)$, provided that the per-tile distributional form (Gaussian, in this work) is sufficiently expressive \citep{ambrogioni2019forward}.\footnote{To investigate whether the marginal posteriors of $\gamma_1^{(b,t)}$, $\gamma_2^{(b,t)}$, and $\kappa^{(b,t)}$ are skewed, heavy-tailed, or multimodal (and thus poorly approximated by a Gaussian distribution) for any bins $b$ or tiles $t$, we also tried approximating them with a nonparametric variational distribution proposed by \cite{mcnamara2026neural}. The fitted densities resembled a Gaussian for most tiles.} Therefore, we expect credible intervals formed using the fitted variational factors to be well-calibrated, and we expect the means of these variational factors to be informative point estimates for shear and convergence in each tile.

Since shear and convergence are derived from the same underlying gravitational potential, one could alternatively parametrize the variational family in terms of just one of these quantities, or in terms of the lensing potential itself. Either alternative would enforce the physical relationship between shear and convergence by construction, but would require additional steps to recover both shear and convergence maps. In the former case, one would need to apply the Kaiser-Squires inversion to samples from the fitted variational distribution \citep{kaiser1993mapping}, while in the latter case, one would need to differentiate samples of the lensing potential. These steps could introduce additional sources of error (e.g., boundary effects from Kaiser-Squires inversion, inflated posterior variance from differentiating the lensing potential), though we have not investigated them empirically.

\subsection{Network architecture} \label{subsec:trainingarchitecture}

Our neural network $f_\eta$ is a residual network (ResNet) of comparable depth to ResNet-34 \citep{he2016deep}. It consists of several residual blocks that progressively downsample the spatial resolution of the input from $2048{\times}2048$ to $8{\times}8$ while simultaneously expanding and compressing the channel depth. The residual blocks are preceded by a preprocessing block and succeeded by a pointwise convolutional layer that maps the final hidden channels to variational parameters $\phi$.

We refined this architecture through a systematic ablation procedure. Starting from a basic ResNet, we iteratively modified the maximum channel depth, number of residual blocks, type of weight normalization, and choice of activation function. Our final architecture, which we selected based on validation loss, has a maximum channel depth of 1024 and 14 residual blocks that utilize group normalization and the SiLU activation function. We illustrate this architecture in \cref{fig:architecturediagram}.

While images are the primary input to the network, we provide the point spread function (PSF) parameters as additional inputs using the approach proposed by \cite{patel2025neural}. The DC2 catalog reports four PSF parameters for each object: three adaptive second moments and the full width at half maximum. For each image, we compute the median of each PSF parameter across all objects measured by the LSST Science Pipelines, and provide these medians to the network as feature maps concatenated with the image. Our ablation study revealed that explicitly providing these PSF feature maps yields only a marginal improvement in the network's average validation loss. This corroborates the insight from \cite{patel2025neural} that exposing the network to images with different PSFs is nearly as beneficial as explicitly supplying the fiducial PSF parameters.

\section{Training and evaluation} \label{sec:traineval}

\subsection{Training the neural network} \label{subsec:training}

We train the network $f_\eta$ on 313 DC2 images and use an additional 39 images for validation. In \cref{sec:results}, we evaluate the accuracy and calibration of the posteriors inferred by the trained network for the remaining 40 images. To minimize information leakage between the three stages of inference, the training, validation, and test sets are spatially disjoint, with each covering a contiguous region.

Training $f_\eta$ takes approximately 16 hours on one NVIDIA A40 GPU. We augment the training set on the fly by applying random 90-degree rotations and vertical flips. The former transformation maps $[\gamma_1^{(b,t)}, \gamma_2^{(b,t)}, \kappa^{(b,t)}] \mapsto [-\gamma_1^{(b,t)}, -\gamma_2^{(b,t)}, \kappa^{(b,t)}]$ for each bin $b$ and tile $t$, while the latter maps $[\gamma_1^{(b,t)}, \gamma_2^{(b,t)}, \kappa^{(b,t)}] \mapsto [\gamma_1^{(b,t)}, -\gamma_2^{(b,t)}, \kappa^{(b,t)}]$. We apply a random combination of these transformations to each training batch, updating the fiducial shear and convergence accordingly before computing the loss.

\begin{figure*}
    \centering
	\includegraphics[width=\textwidth]{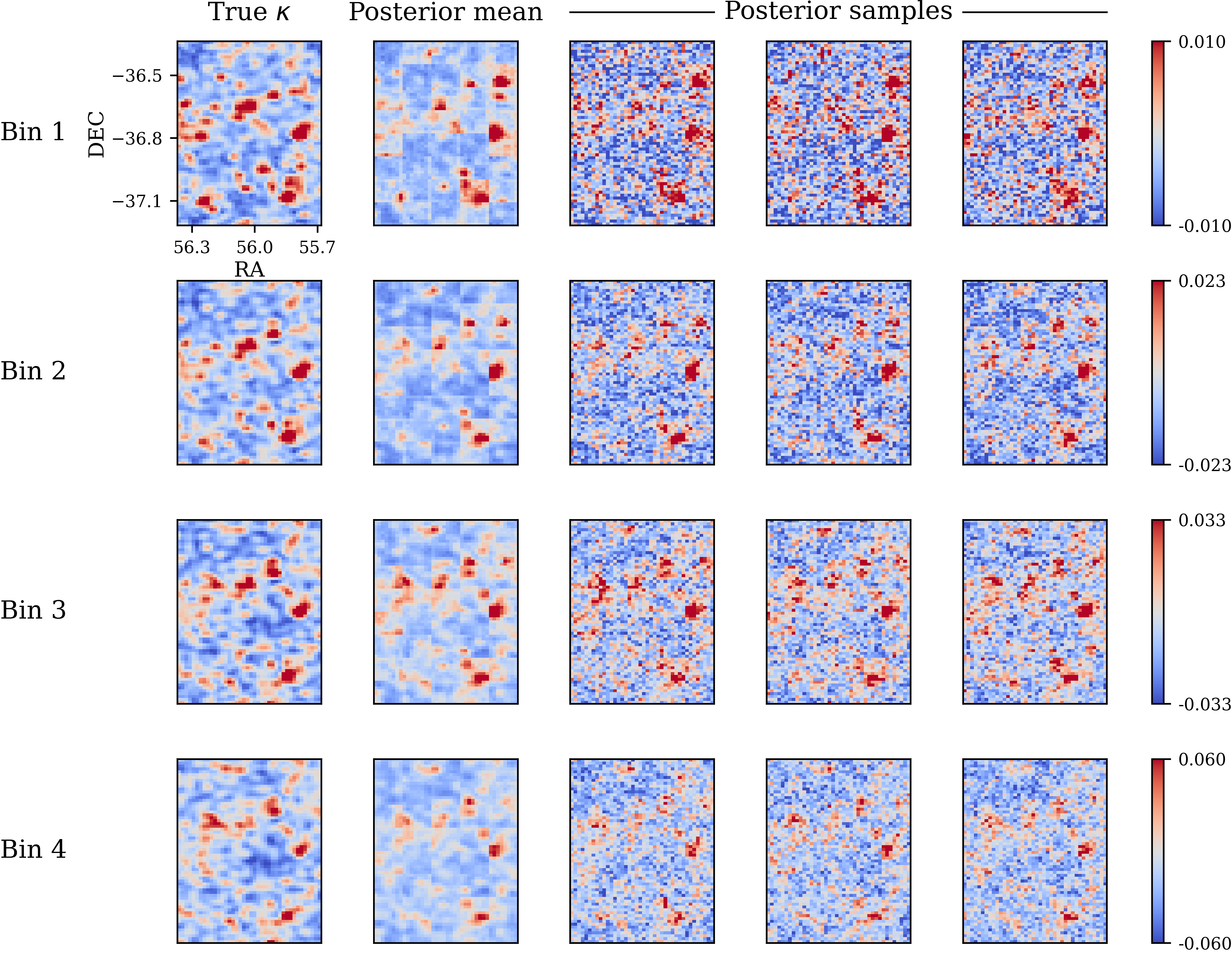}
    \caption{Actual tomographic convergence map for the 0.6 deg$^2$ test set versus the posterior mean and samples from $\prod_{n=1}^N q_{f_\eta(x_n)}(\kappa_n \padvert x_n)$, the variational distribution for convergence across the $N{=}40$ test images. The horizontal and vertical axes are right ascension (RA) and declination (DEC), respectively.}
    \label{fig:dc2convergencemaps}
\end{figure*}

\subsection{Evaluating probabilistic shear and convergence inference} \label{subsec:evaluation}

While there are standard summary statistics and metrics for assessing point estimates produced by deterministic shear and convergence estimators, procedures for evaluating probabilistic weak lensing inference are less established. We adopt a few strategies from the probabilistic mass mapping literature \citep{schneider2017probabilistic, remy2023probabilistic, boruah2024map,boruah2024bayesian}.

First, for each image $x$ in the test set, we plot the mean and several samples from the variational distribution $q_{f_\eta(x)}(z \padvert x)$. Since the images in our test set cover a contiguous region, we also arrange the posterior maps by right ascension (RA) and declination (DEC). This is equivalent to visualizing the mean and samples from $\prod_{n=1}^N q_{f_\eta(x_n)}(z_n \padvert x_n)$, our approximation of the joint posterior over the region spanned by the $N{=}40$ test images.

Second, we evaluate the accuracy of the posterior mean shear (denoted $\widehat{\gamma}_1$, $\widehat{\gamma}_2$) and convergence ($\widehat{\kappa}$) maps as point estimates for $\gamma_1$, $\gamma_2$, and $\kappa$, respectively. For each redshift bin, we compute the root mean squared error (RMSE) and Pearson correlation coefficient ($r$) between the posterior mean and the ground truth across the $NHW$ tiles in the test images. For convergence in bin $b$, the RMSE is
\begin{equation} \label{eq:rmse}
    \text{RMSE}\left(\kappa^{(b)}, \widehat{\kappa}^{(b)}\right) = \sqrt{\frac{1}{NHW} \sum_{n=1}^N \sum_{t=1}^{HW} \left(\kappa_n^{(b,t)} - \widehat{\kappa}_n^{(b,t)}\right)^2}
\end{equation}
and the Pearson correlation coefficient is
\begin{equation} \label{eq:pearson}
    r\left(\kappa^{(b)}, \widehat{\kappa}^{(b)}\right) = \frac{\text{Cov}\left(\kappa^{(b)}, \widehat{\kappa}^{(b)}\right)}{\sqrt{\text{Var}\left(\kappa^{(b)}\right) \text{Var}\left(\widehat{\kappa}^{(b)}\right)}},
\end{equation}
where $\text{Cov}(\cdot,\cdot)$ denotes the sample covariance of the input vectors and $\text{Var}(\cdot)$ denotes the sample variance. The same definitions apply for the two shear components.

However, distilling the variational distribution to a point estimate is reductive. The metrics above offer no insight into the calibration of the variational distribution, which is an important property if maps sampled from it are to be used for downstream cosmological inference.

To assess calibration, we construct marginal highest-density credible intervals based on $q_{f_\eta(x)}(z \padvert x)$ for the shear and convergence in each tile of each redshift bin. We create calibration plots by plotting the intervals against the corresponding ground truth values. After forming credible intervals for several coverage levels, we compute the proportion of intervals that cover the ground truth for each level. If the marginals of $q_{f_\eta(x)}(z \padvert x)$ are well-calibrated approximations of the corresponding marginals of $p(z \padvert x)$, then the empirical coverage should be close to the nominal coverage for each level.

Finally, we investigate the extent to which the inferred maps recover the spatial structure of the underlying lensing fields. For convergence, we compare tomographic auto- and cross-power spectra $C_\kappa^{ij}(\ell)$ computed using posterior samples to those computed using the actual DC2 maps. For each redshift bin, we also compute the cross-correlation coefficient $r(\ell)$ between the actual DC2 maps and those sampled from the variational distribution. This enables us to identify the scales at which the inferred maps accurately reconstruct the truth.

\section{Results} \label{sec:results}

\subsection{Posterior mass maps} \label{subsec:posteriormaps}

\Cref{fig:dc2convergencemaps} compares the actual tomographic convergence map for the test set to the posterior mean and samples inferred with NPE. \Cref{fig:dc2shearmaps} provides a similar comparison for the two shear components.

The posterior mean convergence map identifies most of the overdensities (i.e., high-convergence regions) and underdensities (i.e., low-convergence regions) present in the actual map. It also correctly identifies that the overdensities and underdensities tend to be larger in absolute value at higher redshifts. In these two ways, the posterior mean map reflects the spatial autocorrelation of convergence. \Cref{fig:crosscorrtruth} confirms this quantitatively: the cross-correlation between the inferred and actual convergence maps remains high across a range of multipoles in all four redshift bins, decreasing at higher $\ell$. The tomographic auto- and cross-spectra of posterior samples track those of the ground truth at low multipoles but diverge at higher multipoles (\cref{fig:crosscorrtomographic}).

The posterior mean map provides a smooth approximation of the actual convergence map within each $8{\times}8$-tile region corresponding to a single DC2 image. However, there are spatial discontinuities at many boundaries where we stitched together maps for adjacent images. The number of boundaries could be reduced by training the network on larger images (e.g., $4100{\times}4100$-pixel patches), but doing so is challenging in practice without substantial computing resources.

\Cref{fig:stdevmaps} plots the posterior standard deviations inferred with NPE for the same region. This figure reveals another boundary effect: the standard deviations are systematically larger for the peripheral tiles of each $8{\times}8$-tile map. As a result, gridlines form when these maps are aligned by RA and DEC. This is a natural consequence of the truncation of the network's receptive field for the peripheral tiles. For the interior tiles, the network's inferences are implicitly conditioned on the ellipticities of galaxies in all surrounding directions. For the peripheral tiles, the network has less context, as it cannot ``see'' galaxies in at least one direction.

\subsection{Accuracy of posterior means} \label{subsec:accuracy}

\begin{table}
    \centering
    \begin{tabular}{cccc}
        \toprule
        & & \textbf{RMSE} & \textbf{Pearson} $r$ \\
        \midrule
        \multirow{3}{*}{Bin 1} & $\widehat{\gamma}_1$ & $4.62{\times}10^{-3}$ & 0.25 \\
                               & $\widehat{\gamma}_2$ & $5.08{\times}10^{-3}$ & 0.29 \\
                               & $\widehat{\kappa}$   & $4.29{\times}10^{-3}$ & 0.62 \\
        \midrule
        \multirow{3}{*}{Bin 2} & $\widehat{\gamma}_1$ & $9.01{\times}10^{-3}$ & 0.31 \\
                               & $\widehat{\gamma}_2$ & $7.53{\times}10^{-3}$ & 0.38 \\
                               & $\widehat{\kappa}$   & $8.02{\times}10^{-3}$ & 0.71 \\
        \midrule
        \multirow{3}{*}{Bin 3} & $\widehat{\gamma}_1$ & $1.15{\times}10^{-2}$ & 0.37 \\
                               & $\widehat{\gamma}_2$ & $1.01{\times}10^{-2}$ & 0.40 \\
                               & $\widehat{\kappa}$   & $1.03{\times}10^{-2}$ & 0.74 \\
        \midrule
        \multirow{3}{*}{Bin 4} & $\widehat{\gamma}_1$ & $1.47{\times}10^{-2}$ & 0.34 \\
                               & $\widehat{\gamma}_2$ & $1.46{\times}10^{-2}$ & 0.37 \\
                               & $\widehat{\kappa}$   & $1.59{\times}10^{-2}$ & 0.69 \\
        \bottomrule
    \end{tabular}
    \caption{Root mean squared error (RMSE) and Pearson correlation coefficient of the posterior mean shear ($\widehat{\gamma}_1$, $\widehat{\gamma}_2$) and convergence ($\widehat{\kappa}$) maps inferred with neural posterior estimation. For each redshift bin, the two metrics are computed using all 2,560 test set tiles (\cref{eq:rmse,eq:pearson}).}
    \label{table:posteriormeanmetrics}
\end{table}

\Cref{table:posteriormeanmetrics} reports the RMSE and Pearson correlation coefficient of the posterior means $\widehat{\gamma}_1$, $\widehat{\gamma}_2$, and $\widehat{\kappa}$ for each redshift bin. These metrics are similar in each bin for $\widehat{\gamma}_1$ and $\widehat{\gamma}_2$, which suggests that our neural network is not rotationally biased.

For all three lensing variables, the RMSE is smallest for the lowest redshift bin and largest for the highest redshift bin. This is consistent with the positive relationship between redshift and amplitude identified in the previous section. The Pearson correlation coefficients are also smallest in the low redshift bins. This suggests that the network struggles to distinguish the weak lensing signal from the shape noise of low-redshift galaxies, even though it can achieve a small RMSE in this setting by inferring that shear and convergence are close to zero.

In each redshift bin, the Pearson correlation coefficient is larger for convergence than for shear, and it is similar for the two shear components. The same trends hold for Spearman and Kendall correlation coefficients, so we elect not to include these rank-based metrics in the table.

\subsection{Posterior calibration} \label{subsec:calibration}

\begin{figure*}
    \begin{tikzpicture}
        \node (credibleintervals) {\includegraphics[width=0.425\textwidth]{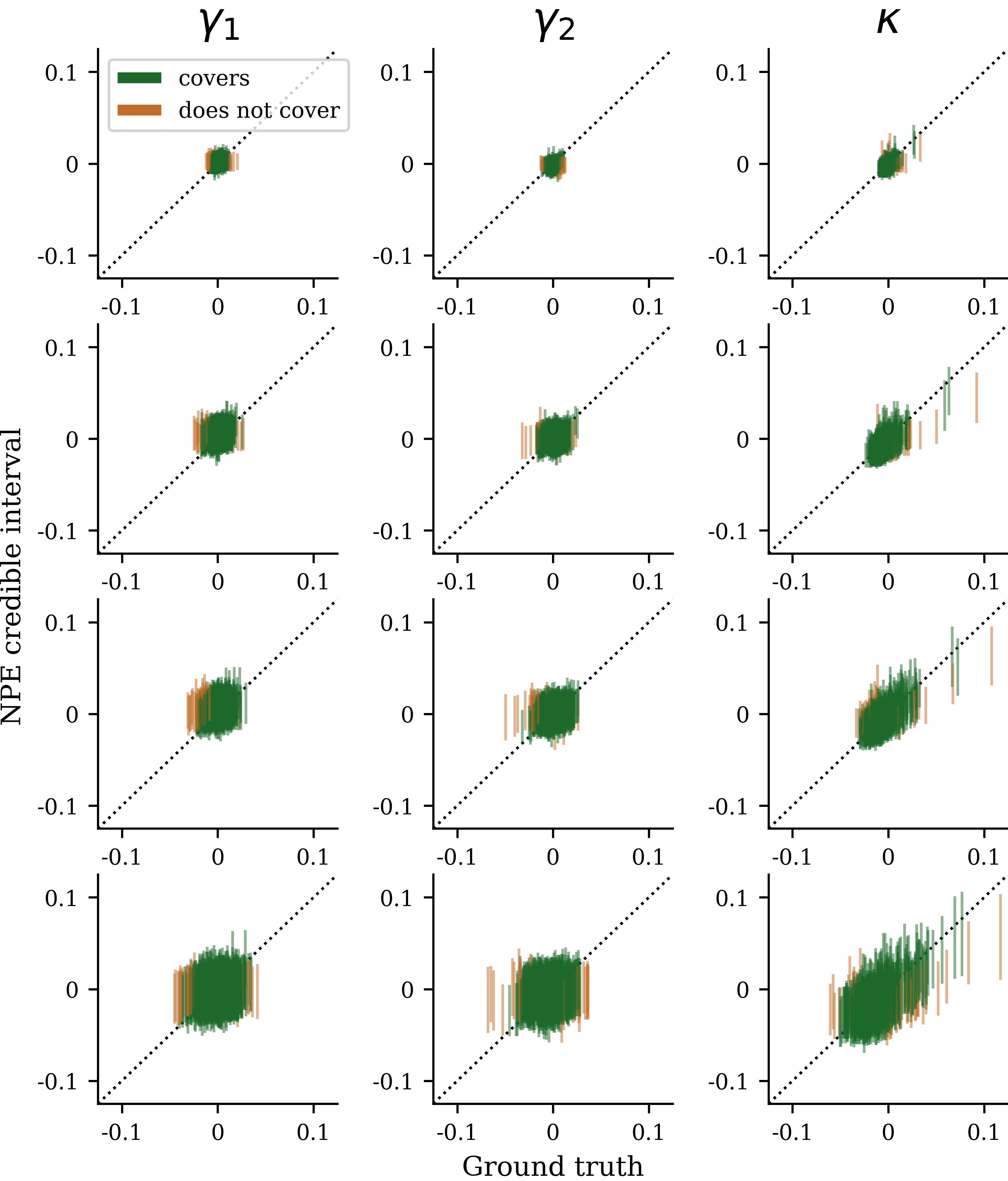}};
        \node (coverageprobs) [right=of credibleintervals, xshift=-10pt] {\includegraphics[width=0.425\textwidth]{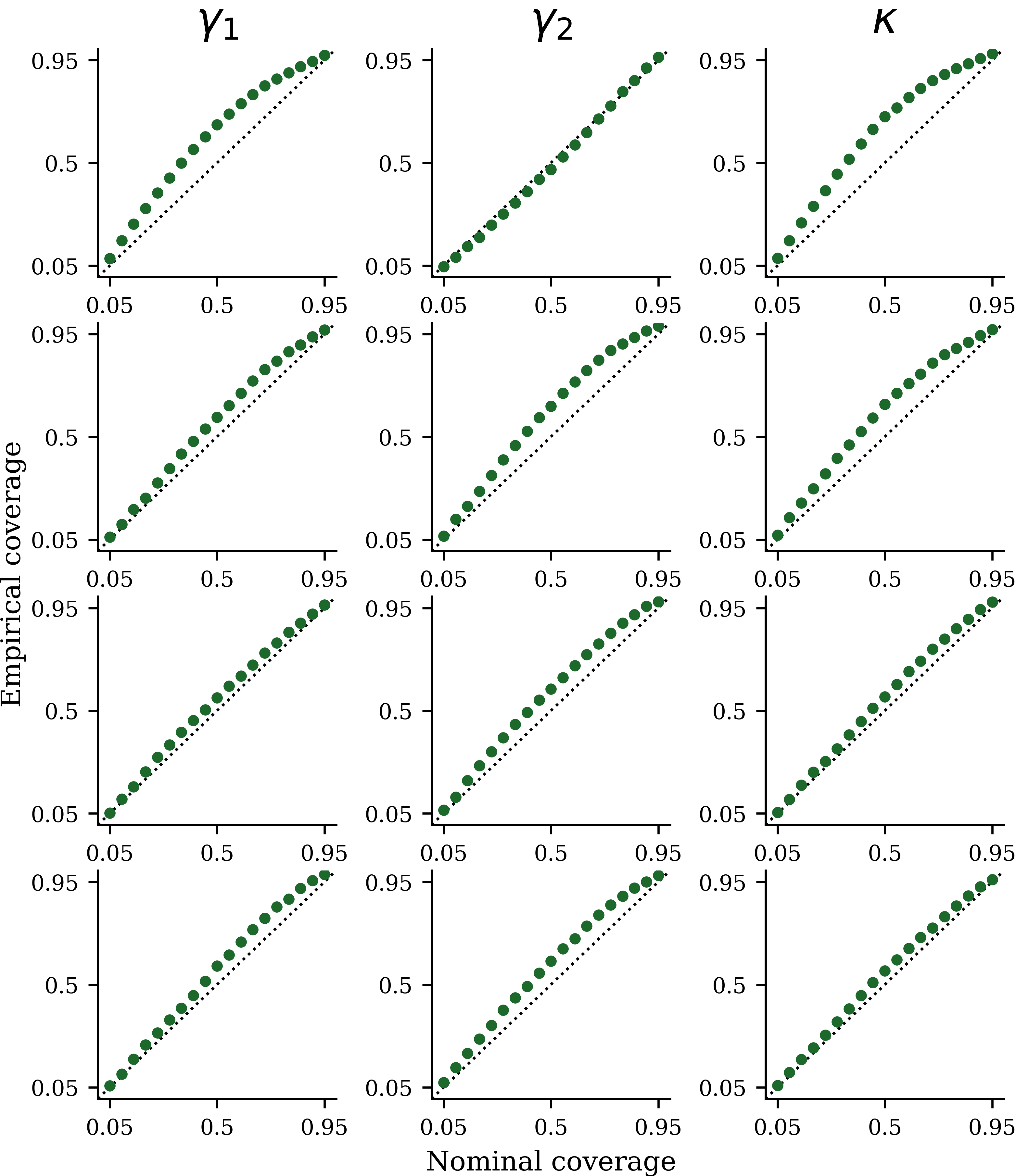}};
        \node (bin1label) [left=of credibleintervals, xshift=20pt, yshift=31pt] {\large Bin 2};
        \node [above=42pt of bin1label] {\large Bin 1};
        \node (bin2label) [below=42pt of bin1label] {\large Bin 3};
        \node [below=42pt of bin2label] {\large Bin 4};
    \end{tikzpicture}
    \caption{Left: 90\% credible intervals for shear and convergence, constructed independently for a random sample of 500 test set tiles using the 0.05 and 0.95 quantiles of the variational distribution. Right: Nominal and empirical coverage of credible intervals for shear and convergence. For this panel, credible intervals were constructed independently for all 2,560 test set tiles.}
    \label{fig:dc2credibleintervals}
\end{figure*}
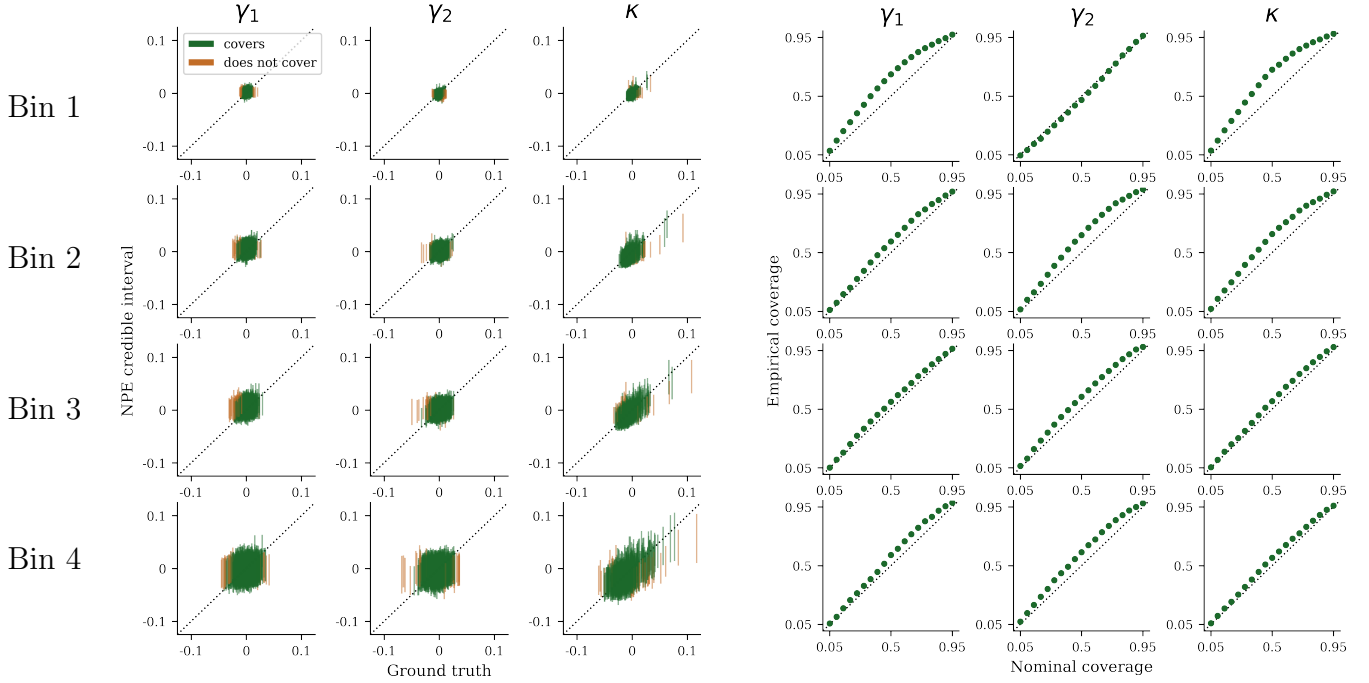

The left panel of \cref{fig:dc2credibleintervals} is a calibration plot that compares the actual shear and convergence for a random sample of test set tiles to 90\% marginal credible intervals formed using $q_{f_\eta(x)}(z \padvert x)$. We represent the intervals as vertical line segments and color them based on whether they cover the ground truth (i.e., intersect the 45-degree line). Approximately 90\% of the intervals cover the ground truth for each lensing variable in each redshift bin.

The credible intervals do not adhere perfectly to the 45-degree line, as they would if the posterior median shear and convergence (i.e., the centers of the intervals) coincided exactly with the actual shear and convergence in each tile. However, the intervals generally slope upward, consistent with the Pearson correlation coefficients reported in the previous section.

Within each redshift bin, the interval widths are approximately uniform across the range of actual shear and convergence values. However, the intervals are much wider for the higher redshift bins. This is consistent with the posterior standard deviation maps in \cref{fig:stdevmaps}.

We repeat the above process of constructing credible intervals and computing empirical coverage for nominal coverage levels ranging from 0.05 to 0.95. The right panel of \cref{fig:dc2credibleintervals} reveals that the empirical coverage is close to the nominal coverage for each lensing variable in each redshift bin, which suggests that the marginal variational distributions inferred with NPE are well-calibrated. Across all twelve subplots, the empirical coverages are slightly larger than the nominal coverages for most levels — i.e., the credible intervals tend to be too wide. This tendency is typical of methods that minimize the forward Kullback-Leibler divergence and is generally considered preferable to the opposite case, where credible intervals are systematically too narrow.

\section{Discussion} \label{sec:discussion}

To our knowledge, this work is the first to investigate the capabilities of NPE as a field-level weak lensing inference method. Our results are promising: when evaluated on a held-out subregion of the DC2 Simulated Sky Survey, NPE infers well-calibrated variational distributions over shear and convergence maps in four redshift bins. For all lensing variables in all bins, the means of the variational distributions are positively correlated with the ground truth across the test set, and marginal credible intervals formed using the variational quantiles cover the ground truth at approximately nominal rates.

\subsection{Remedies for simulator misspecification} \label{subsec:misspecification}

Our assessment of NPE did not address the possibility of distribution shift between the training and evaluation images — nor did it need to, since both sets of images were generated by the DC2 simulator in our experiments. In a more practical scenario, the network would be trained on simulated images and applied to real images for which the underlying lensing maps are unknown. If there are discrepancies between the simulator's implicit generative model and reality in this setting, NPE's posterior approximations could be systematically biased or miscalibrated \citep{hermans2022crisis}. In the context of DC2, this misspecification could occur at any stage of the generative model illustrated in the left panel of \cref{fig:inferencediagram}.

A better understanding of the sensitivity of our NPE procedure to various types of simulator misspecification will be necessary before we can reliably deploy it on Stage IV surveys. Several recent works have proposed strategies to mitigate simulator misspecification \citep{kelly2025simulation}. These strategies typically involve penalizing discrepancies between simulated and real observations \citep{huang2023learning, gloeckler2023adversarial, swierc2024domain, mishra2025robust, elsemuller2025does}, correcting such discrepancies using a calibration set of real observations \citep{wehenkel2024addressing, senouf2025inductive}, or relating simulated and real observations with an explicit error model \citep{ward2022robust, ocallaghan2025misspecification}.

\subsection{A more flexible variational family} \label{subsec:flowmatching}

We evaluated NPE only for a particular mean-field Gaussian variational family. The marginal variational factors of this family were well-calibrated, and maps sampled from the fitted variational distribution recovered the long-range spatial dependencies of the underlying shear and convergence fields. However, because our variational family factorizes across tiles, redshift bins, and lensing variables, the posterior mass maps did not exhibit auto- or cross-correlation at the tile level.

A natural next step is therefore to replace the mean-field family with a more expressive parametrization. Conditional flow matching could be used to fit a more flexible variational family that captures the auto- and cross-correlation of shear and convergence in the exact posterior \citep{lipman2022flow, wildberger2023flow}.

Maps sampled using a trained flow would provide a smooth approximation of the shear and convergence fields within $8{\times}8$-tile regions, assuming the same map resolution used in this work. However, the boundary effects observed in \cref{fig:dc2convergencemaps} and \cref{fig:stdevmaps} would persist. A potential remedy for these boundary effects would be to use an autoregressive tiling scheme similar to that proposed by \cite{regier2026neural}.

\subsection{A path to posterior inference of cosmological parameters} \label{subsec:cosmologicalparams}

The ultimate objective of weak lensing analyses is to constrain cosmological parameters, typically those from the $\Lambda$CDM model or one of its variants. Training NPE to directly map survey-scale images to cosmological parameters would be computationally prohibitive, as it would require simulating massive synthetic sky surveys for each plausible combination of cosmological parameters. A more tractable alternative is to decompose the problem hierarchically.

Let $\theta$ denote cosmological parameters of interest, and let $\kappa$ denote the convergence map for a region of interest $x$, which could be the survey footprint or a large contiguous subregion of it. This overloads our earlier notation, where $\kappa$ referred to a single map for a single image $x$.

Under this formulation, the posterior distribution over cosmological parameters can be written as
\begin{align}
    p(\theta \padvert x) &= \int p(\theta \padvert \kappa, x) p(\kappa \padvert x) d\kappa \approx \int p(\theta \padvert \kappa) p(\kappa \padvert x) d\kappa,
\end{align}
where the approximation reflects the assumption that the convergence field captures most of the cosmological information encoded in the images. Although this conditional independence is not exact, it is a reasonable approximation when $\theta$ comprises parameters primarily constrained by weak lensing — namely, $\Omega_m$ and $\sigma_8$. This perspective aligns with the standard practice of treating convergence maps (or summary statistics derived from them) as approximately sufficient statistics for cosmological parameter inference.

Approximating $p(\theta \padvert x)$ via this decomposition requires two subroutines. The first infers $q_\phi(\kappa \padvert x)$ using the NPE procedure described in this work, but with a variational family flexible enough to capture the spatial autocorrelation of $\kappa$ in the exact posterior (\cref{subsec:flowmatching}). The second infers $q_\psi(\theta \padvert \kappa)$ using NPE or another neural simulation-based inference algorithm trained on $(\theta, \kappa)$ pairs generated by a high-fidelity simulator \citep{zeghal2024simulation}. Given the networks trained in these subroutines, one could approximate
\begin{align}
    q(\theta \padvert x) &\approx \int q_\psi(\theta \padvert \kappa) q_\phi(\kappa \padvert x) d\kappa
\end{align}
via ancestral sampling, by first drawing convergence maps from $q_\phi(\kappa \padvert x)$ and then passing them through the second-stage network to obtain samples from $q_\psi(\theta \padvert \kappa)$. Credible regions for the parameters $\theta$ could then be constructed from these samples.

This decomposition would enable end-to-end probabilistic weak lensing inference at the scale of Stage IV astronomical surveys. The present work establishes the feasibility of the images-to-maps stage of this pipeline and proposes the extensions required for it to produce trustworthy inferences on real survey data.

\begin{acknowledgments}
This material is based on work supported by the National Science Foundation under Grant No. 2209720 and the U.S. Department of Energy, Office of Science, Office of High Energy Physics under Award Number DE-SC0023714. The authors thank Steve Fan and Tahseen Younus for helpful discussions about this work and for their contributions to the blissWL codebase.

This paper has undergone internal review in the LSST Dark Energy Science Collaboration. The internal reviewers were Supranta Boruah, Alan Heavens, and Benjamin Remy. The authors thank Pat Burchat for helpful comments during the collaboration-wide review period.

The DESC acknowledges ongoing support from the Institut National de Physique Nucl\'eaire et de Physique des Particules in France; the Science \& Technology Facilities Council in the United Kingdom; and the Department of Energy and the LSST Discovery Alliance in the United States. DESC uses resources of the IN2P3 Computing Center (CC-IN2P3--Lyon/Villeurbanne - France) funded by the Centre National de la Recherche Scientifique; the National Energy Research Scientific Computing Center, a DOE Office of Science User Facility supported by the Office of Science of the U.S.\ Department of Energy under Contract No.\ DE-AC02-05CH11231; STFC DiRAC HPC Facilities, funded by UK BEIS National E-infrastructure capital grants; and the UK particle physics grid, supported by the GridPP Collaboration. This work was performed in part under DOE Contract DE-AC02-76SF00515.
\end{acknowledgments}

\begin{contribution}
TW, SC, and JR extended the blissWL software package to infer DC2 mass maps with NPE. TW trained and evaluated the neural network and created the figures. CA and JR supervised the project. TW, CA, and JR wrote the text of this manuscript.
\end{contribution}

\section*{Data availability}

Our implementation of NPE is publicly available at \url{https://github.com/prob-ml/blissWL}. Code for reproducing the tables and figures in this paper can be found in the \texttt{images\_to\_maps/dc2} folder.



\bibliographystyle{aasjournalv7}
\bibliography{references}



\clearpage
\appendix
\crefalias{section}{appendix}
\setcounter{figure}{0}
\setcounter{table}{0}
\renewcommand{\thefigure}{\thesection\arabic{figure}}
\renewcommand{\thetable}{\thesection\arabic{table}}
\makeatletter
\@addtoreset{figure}{section}
\@addtoreset{table}{section}
\makeatother

\section{Additional figures} \label{appdx:additionalfigs}

\begin{minipage}{\textwidth}
    \centering
    \includegraphics[width=0.4\textwidth]{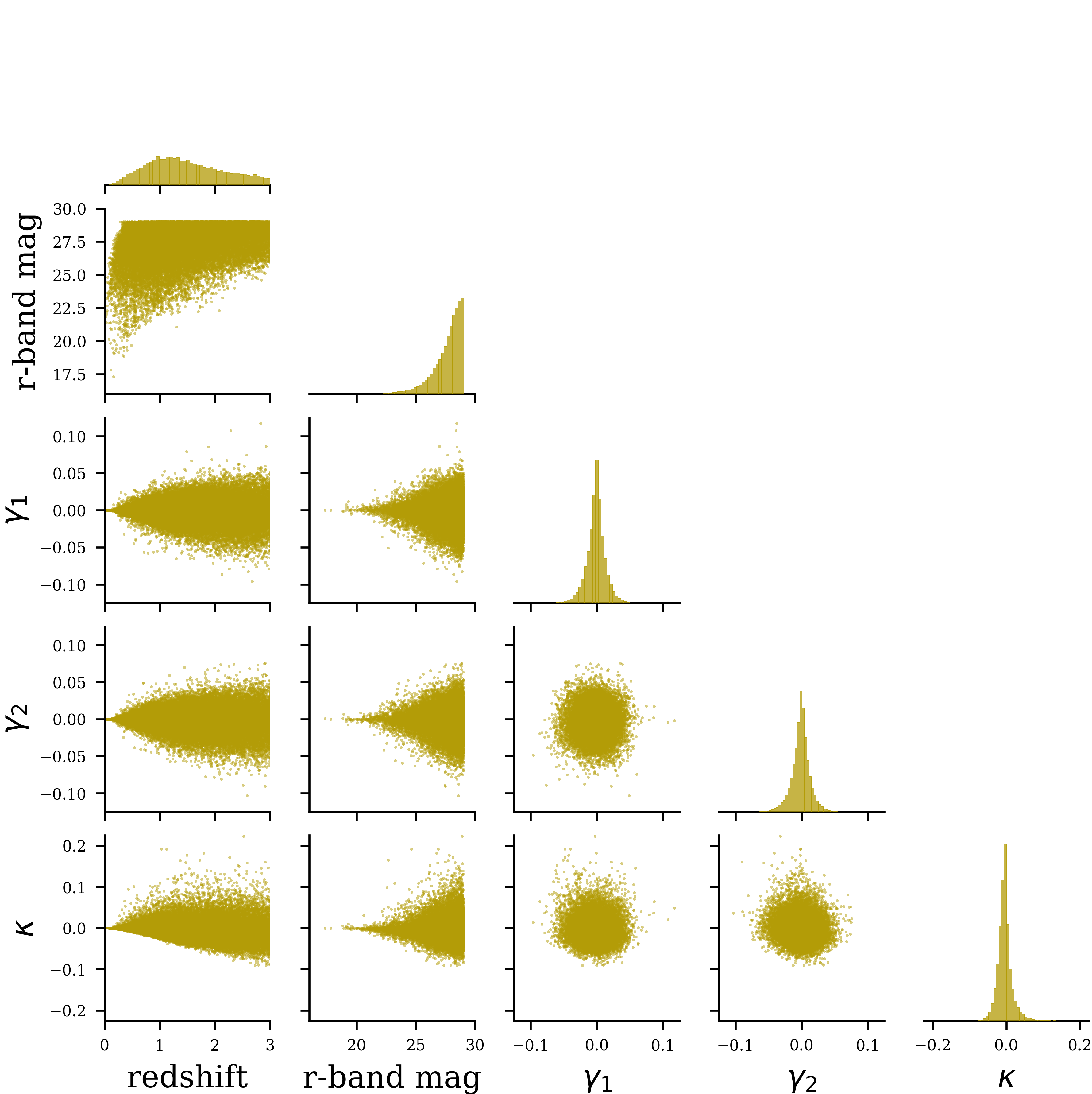}
    \captionof{figure}{Marginal histograms and pairwise scatter plots of redshift, $r$-band magnitude, shear, and convergence for 50,000 galaxies whose centroids lie in tract 3828 or 3829 of the DC2 Simulated Sky Survey.}
    \label{fig:dc2galaxyproperties}
\end{minipage}

\vfill

\begin{minipage}{\textwidth}
    \centering
	\includegraphics[width=0.7\textwidth]{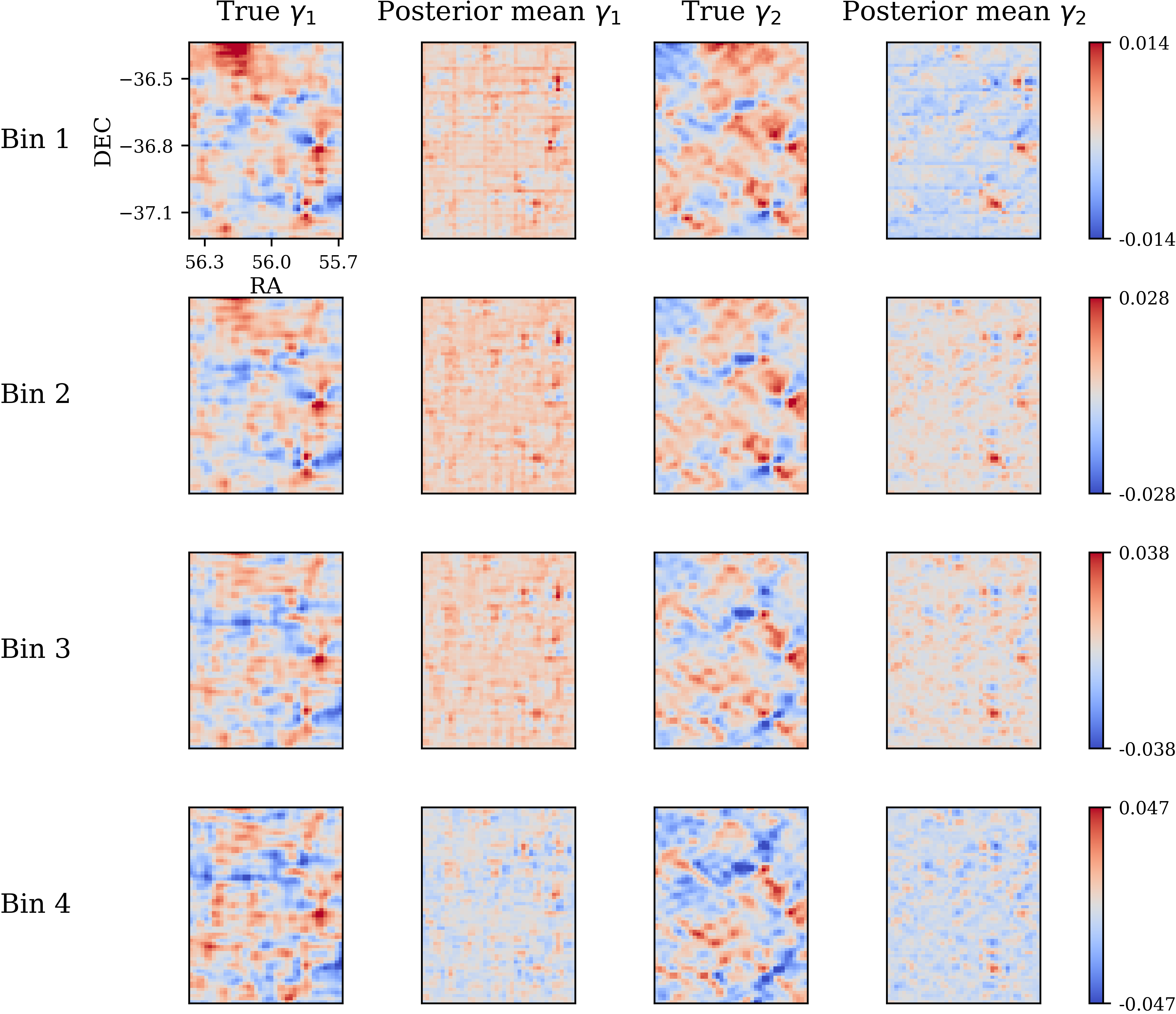}
    \captionof{figure}{Actual tomographic shear maps for the 0.6 deg$^2$ test set versus the posterior mean of $\prod_{n=1}^N q_{f_\eta(x_n)}(\gamma_{1n}, \gamma_{2n} \padvert x_n)$, the variational distribution for shear across the $N{=}40$ test images. Samples from this variational distribution are of poor quality for the reasons discussed in \cref{subsec:variationalfamily}, so we omit them here. The horizontal and vertical axes are right ascension (RA) and declination (DEC), respectively.}
    \label{fig:dc2shearmaps}
\end{minipage}

\begin{minipage}{\textwidth}
    \centering
	\includegraphics[width=0.5\textwidth]{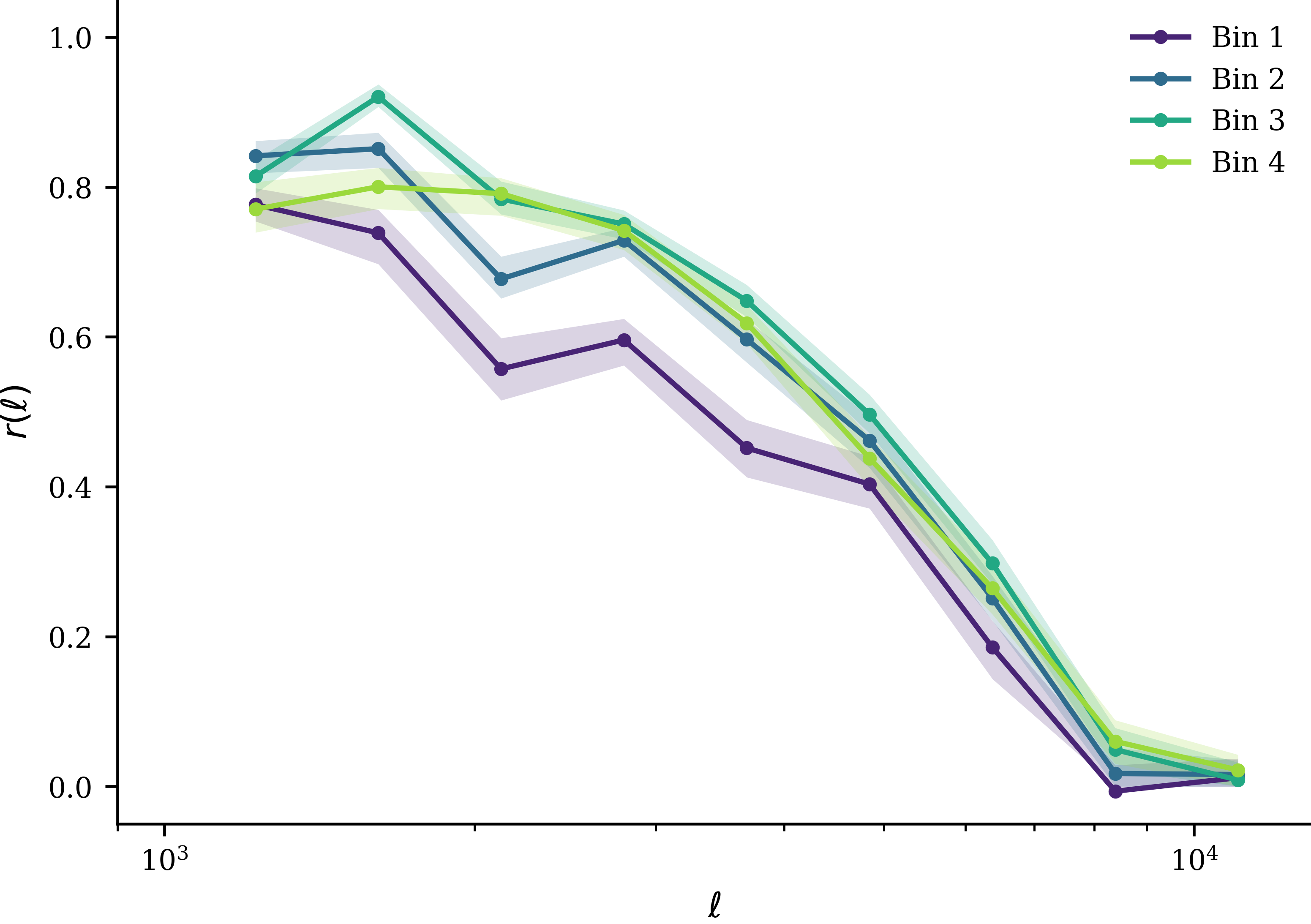}
    \captionof{figure}{Cross-correlation coefficient $r(\ell) = C_{tp}(\ell) / \sqrt{C_{tt}(\ell) C_{pp}(\ell)}$ between the actual and posterior median convergence maps as a function of multipole $\ell$, in each of the four tomographic redshift bins. The solid lines represent the median across 200 samples drawn from the variational distributions inferred by NPE, while the shaded bands span from the 0.25 quantile to the 0.75 quantile of these samples. Spectra are computed via 2D FFT of the stitched $64 \times 40$ tile test-set map and radially binned in nine log-spaced $\ell$ bins.}
    \label{fig:crosscorrtruth}
\end{minipage}

\begin{minipage}{\textwidth}
    \centering
	\includegraphics[width=0.7\textwidth]{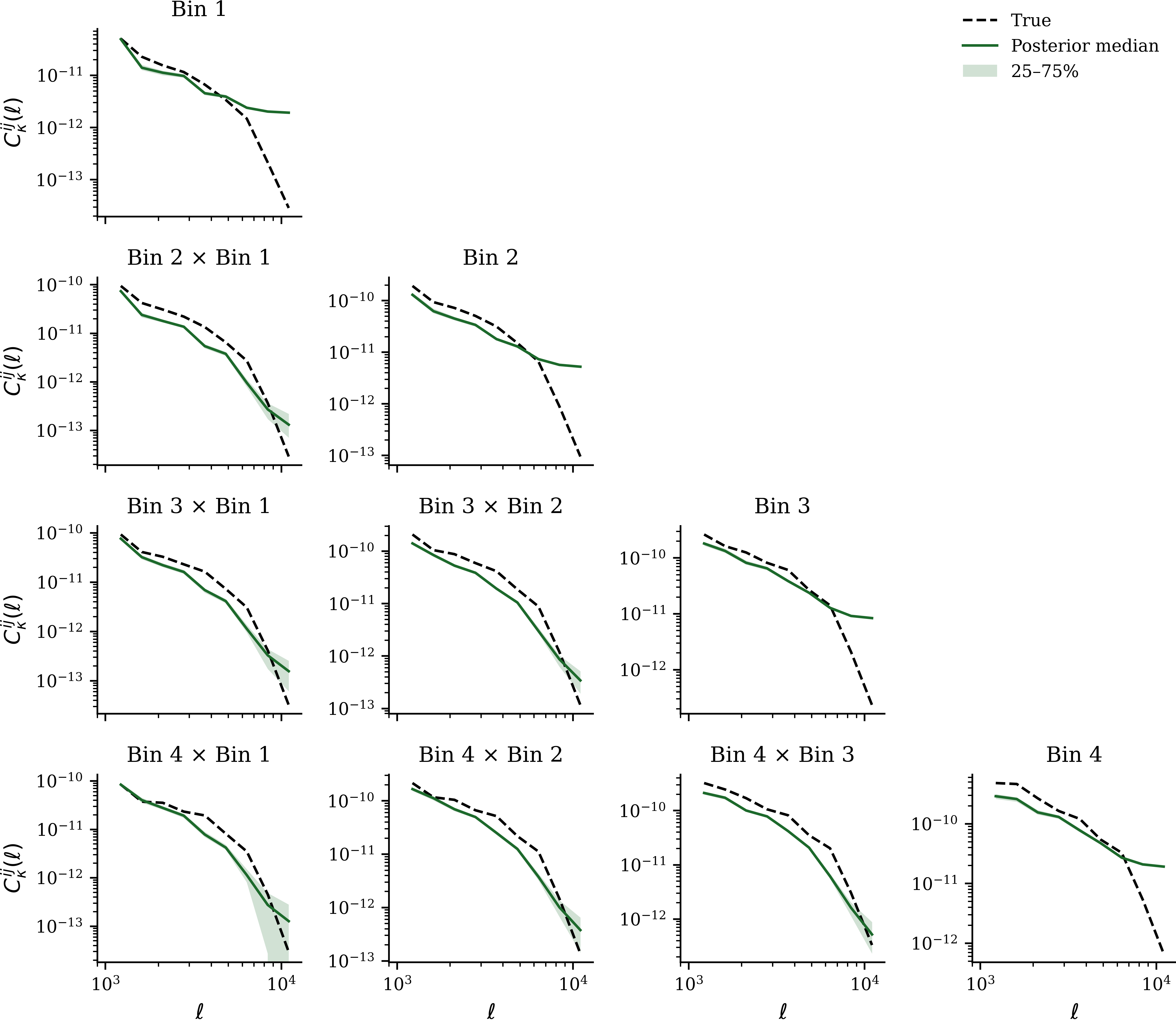}
    \captionof{figure}{Auto- and cross-power spectra of the convergence field, $C_\kappa^{ij}(\ell)$, for the four redshift bins ($i, j \in \{1, 2, 3, 4\}$). The diagonal panels display the auto-power spectra, while the off-diagonal panels display the cross-power spectra. Black dashed curves are the power spectra computed using the actual DC2 convergence maps from the 40-image test set, stitched into a single $64 \times 40$ tile field covering approximately 0.6 deg$^2$. The green curves are the medians across 200 samples drawn from the variational distributions inferred by NPE. Each shaded band spans from the 0.25 quantile to the 0.75 quantile of the 200 samples. Spectra are estimated by 2D FFT of the stitched maps and radially averaged in nine log-spaced $\ell$ bins.}
    \label{fig:crosscorrtomographic}
\end{minipage}

\begin{minipage}{\textwidth}
    \centering
	\includegraphics[width=0.7\textwidth]{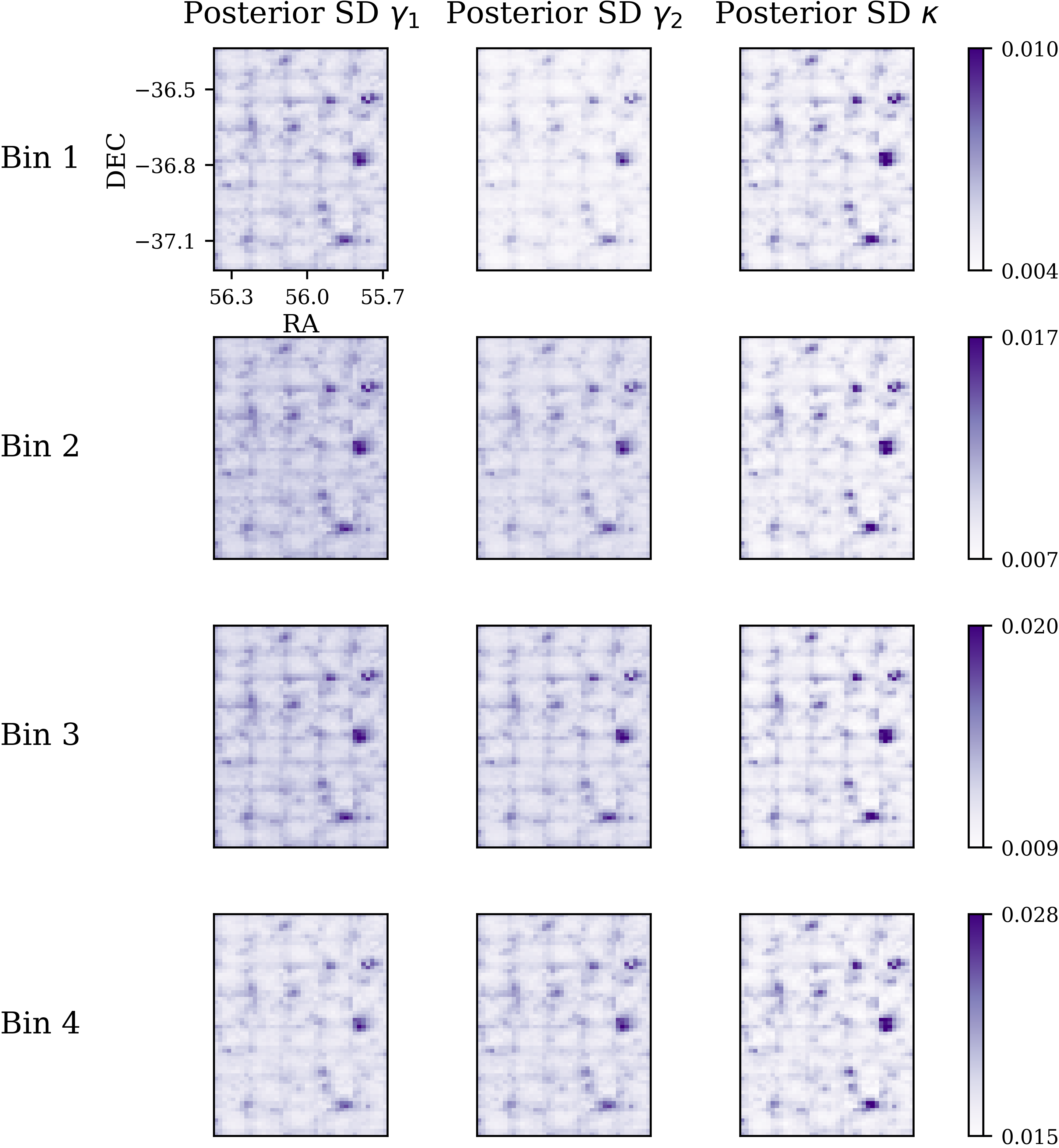}
    \captionof{figure}{Tomographic standard deviation maps for the 0.6 deg$^2$ test set. These maps depict the marginal standard deviations of $\prod_{n=1}^N q_{f_\eta(x_n)}(z_n \padvert x_n)$, the variational distribution for shear and convergence across the $N{=}40$ test images. The horizontal and vertical axes are right ascension (RA) and declination (DEC), respectively.}
    \label{fig:stdevmaps}
\end{minipage}

\vfill

\section{Neural network architecture} \label{appdx:architecture}

\Cref{fig:architecturediagram} illustrates the architecture of our neural network $f_\eta$, which maps a PSF-augmented image $x$ to variational parameters $\phi$. The network's inputs and outputs (outlined in black) are connected by a series of individual layers (shaded gray) and multi-layer residual blocks (shaded blue).

The backbone of the network consists of 14 residual blocks. Each residual block transforms the network's intermediate representations based on an input channel count $c$, an output channel count $c^\prime$, and a stride. The resulting channel depth and spatial resolution of these intermediate representations are listed in italics on the left side of the diagram. While these tensors also have a batch dimension, we omit it from the diagram. As explained in \cref{subsec:npeobjective}, we use a batch size of one to accommodate GPU memory constraints.

The panel on the right side of the diagram details the internal structure of a single residual block. Each block employs two 3x3 convolutional layers interspersed with group normalization and SiLU activation layers \citep{wu2018group}, as well as a skip connection composed of a 1x1 convolution and group normalization.

\clearpage

\begin{minipage}{\textwidth}
    \centering
    \resizebox{\textwidth}{!}{%
    \begin{tikzpicture}[
        inputoutput/.style={
        rectangle, ultra thick, draw=black, align=center, minimum width=1in
        },
        resblockoutline/.style={
        rectangle, ultra thick, dotted, rounded corners, draw=mediumteal, align=center, minimum width=2in
        },
        resblock/.style={
        rectangle, thick, rounded corners, draw=mediumteal, fill=mediumteal!10, align=center, minimum width=2in
        },
        preprocesslayer/.style={
        rectangle, thick, rounded corners, draw=pinkgray, fill=pinkgray!10, align=center, minimum width=1.25in
        },
        resblocklayer/.style={
        rectangle, thick, rounded corners, draw=pinkgray, fill=pinkgray!10, align=center, minimum width=1.25in
        },
        finallayer/.style={
        rectangle, thick, rounded corners, draw=pinkgray, fill=pinkgray!10, align=center, minimum width=1.25in
        },
        dimensions/.style={
        align=center, font=\itshape
        },
        arrow/.style={
        ->, thick, shorten <=0.1cm, shorten >=0.1cm
        },
    ]
        \node[resblock, minimum width = 2.75in] (resblocktitle) {ResidualBlock(c, c$^\prime$, s)};
        \node[inputoutput, below=0.15in of resblocktitle, xshift=-0.7in] (in) {Input\\
        c $\times$ h $\times$ w};
        \node[resblocklayer, below=0.15in of in] (conv1) {3x3 Conv\\
        c$_{\text{in}}$ = c, c$_{\text{out}}$ = c$^\prime$, stride = s};
        \node[resblocklayer, below=0.15in of conv1] (gn1) {GroupNorm};
        \node[resblocklayer, right=0.15in of gn1, yshift=-0.2in] (outerconv) {1x1 Conv\\
        c$_{\text{in}}$ = c, c$_{\text{out}}$ = c$^\prime$, stride = s};
        \node[resblocklayer, below=0.15in of outerconv] (outergn) {GroupNorm};
        \node[resblocklayer, below=0.15in of gn1] (silu1) {SiLU};
        \node[resblocklayer, below=0.15in of silu1] (conv2) {3x3 Conv\\
        c$_{\text{in}}$ = c$^\prime$, c$_{\text{out}}$ = c$^\prime$, stride = 1};
        \node[resblocklayer, below=0.15in of conv2] (gn2) {GroupNorm};
        \node[circle, thick, draw=gray, below=0.15in of gn2] (connection) {+};
        \node[resblocklayer, below=0.15in of connection] (silu2) {SiLU};
        \node[inputoutput, below=0.15in of silu2] (out) {Output\\
        c$^\prime$ $\times$ $\frac{\text{h}}{\text{s}}$ $\times$ $\frac{\text{w}}{\text{s}}$};
        \node[resblockoutline, fit={(resblocktitle) (in) (conv1) (gn1) (outerconv) (outergn) (silu1) (conv2) (gn2) (silu2) (connection) (out)}] (resblocktemplate) {};

        \draw[arrow, out=0, in=90] (in.east) to (outerconv.north);
        \draw[arrow] (outerconv.south) to (outergn.north);
        \draw[arrow, out=270, in=0] (outergn.south) to (connection.east);
        \draw[arrow] (in.south) to (conv1.north);
        \draw[arrow] (conv1.south) to (gn1.north);
        \draw[arrow] (gn1.south) to (silu1.north);
        \draw[arrow] (silu1.south) to (conv2.north);
        \draw[arrow] (conv2.south) to (gn2.north);
        \draw[arrow] (gn2.south) to (connection.north);
        \draw[arrow] (connection.south) to (silu2.north);
        \draw[arrow] (silu2.south) to (out.north);

        \node[resblock, left=0.5in of resblocktemplate] (rb3) {ResidualBlock(c=128, c$^\prime$=128, stride=1)};
        \node[resblock, above=0.25in of rb3] (rb2) {ResidualBlock(c=64, c$^\prime$=128, stride=2)};
        \node[resblock, above=0.25in of rb2] (rb0) {ResidualBlock(c=64, c$^\prime$=64, stride=1) $\times$ 2};
        \node[resblock, below=0.25in of rb3] (rb4) {ResidualBlock(c=128, c$^\prime$=256, stride=2)};
        \node[resblock, below=0.25in of rb4] (rb5) {ResidualBlock(c=256, c$^\prime$=256, stride=1)};
        \node[resblock, below=0.25in of rb5] (rb6) {ResidualBlock(c=256, c$^\prime$=512, stride=2)};
        \node[resblock, below=0.25in of rb6] (rb7) {ResidualBlock(c=512, c$^\prime$=512, stride=1)};
        \node[resblock, below=0.25in of rb7] (rb8) {ResidualBlock(c=512, c$^\prime$=1024, stride=2)};
        \node[resblock, below=0.25in of rb8] (rb9) {ResidualBlock(c=1024, c$^\prime$=1024, stride=1)};
        \node[resblock, below=0.25in of rb9] (rb10) {ResidualBlock(c=1024, c$^\prime$=1024, stride=2)};
        \node[resblock, below=0.25in of rb10] (rb11) {ResidualBlock(c=1024, c$^\prime$=512, stride=2)};
        \node[resblock, below=0.25in of rb11] (rb12) {ResidualBlock(c=512, c$^\prime$=256, stride=2)};
        \node[resblock, below=0.25in of rb12] (rb13) {ResidualBlock(c=256, c$^\prime$=128, stride=2)};

        \draw[thick, teal, dotted] (rb2.east) to (resblocktemplate.north west);
        \draw[thick, teal, dotted] (rb2.east) to (resblocktemplate.south west);
        
        \draw[arrow] (rb0.south) to node[midway, left=1.25in, dimensions] {64 $\times$ 2048 $\times$ 2048} (rb2.north);
        \draw[arrow] (rb2.south) to node[midway, left=1.25in, dimensions] {128 $\times$ 1024 $\times$ 1024} (rb3.north);
        \draw[arrow] (rb3.south) to node[midway, left=1.25in, dimensions] {128 $\times$ 1024 $\times$ 1024} (rb4.north);
        \draw[arrow] (rb4.south) to node[midway, left=1.25in, dimensions] {256 $\times$ 512 $\times$ 512} (rb5.north);
        \draw[arrow] (rb5.south) to node[midway, left=1.25in, dimensions] {256 $\times$ 512 $\times$ 512} (rb6.north);
        \draw[arrow] (rb6.south) to node[midway, left=1.25in, dimensions] {512 $\times$ 256 $\times$ 256} (rb7.north);
        \draw[arrow] (rb7.south) to node[midway, left=1.25in, dimensions] {512 $\times$ 256 $\times$ 256} (rb8.north);
        \draw[arrow] (rb8.south) to node[midway, left=1.25in, dimensions] {1024 $\times$ 128 $\times$ 128} (rb9.north);
        \draw[arrow] (rb9.south) to node[midway, left=1.25in, dimensions] {1024 $\times$ 128 $\times$ 128} (rb10.north);
        \draw[arrow] (rb10.south) to node[midway, left=1.25in, dimensions] {1024 $\times$ 64 $\times$ 64} (rb11.north);
        \draw[arrow] (rb11.south) to node[midway, left=1.25in, dimensions] {512 $\times$ 32 $\times$ 32} (rb12.north);
        \draw[arrow] (rb12.south) to node[midway, left=1.25in, dimensions] {256 $\times$ 16 $\times$ 16} (rb13.north);
        
        \node[preprocesslayer, above=0.25in of rb0] (preprocess_silu) {SiLU};
        \node[preprocesslayer, above=0.25in of preprocess_silu] (preprocess_gn) {GroupNorm};
        \node[preprocesslayer, above=0.25in of preprocess_gn] (preprocess_conv) {5x5 Conv3d\\
        c$_{\text{in}}$ = 6, c$_{\text{out}}$ = 64, stride = 1};
        \node[inputoutput, above=0.25in of preprocess_conv, align=center] (images) {Image \& PSF feature maps $x$};
        
        \draw[arrow] (images.south) to node[midway, left=1.25in, dimensions] {6 $\times$ 5 $\times$ 2048 $\times$ 2048} (preprocess_conv.north);
        \draw[arrow] (preprocess_conv.south) to node[midway, left=1.25in, dimensions] {64 $\times$ 2048 $\times$ 2048} (preprocess_gn.north);
        \draw[arrow] (preprocess_gn.south) to node[midway, left=1.25in, dimensions] {64 $\times$ 2048 $\times$ 2048} (preprocess_silu.north);
        \draw[arrow] (preprocess_silu.south) to node[midway, left=1.25in, dimensions] {64 $\times$ 2048 $\times$ 2048} (rb0.north);

        \node[finallayer, below=0.25in of rb13] (finalconv) {1x1 Conv\\
        c$_{\text{in}}$ = 128, c$_{\text{out}}$ = 24, stride = 1};
        \node[inputoutput, below=0.25in of finalconv] (varparams) {Variational parameters $\phi$};
        
        \draw[arrow] (rb13.south) to node[midway, left=1.25in, dimensions] {128 $\times$ 8 $\times$ 8} (finalconv.north);
        \draw[arrow] (finalconv.south) to node[midway, left=1.25in, dimensions] {24 $\times$ 8 $\times$ 8} (varparams.north);
    \end{tikzpicture}
    }
    
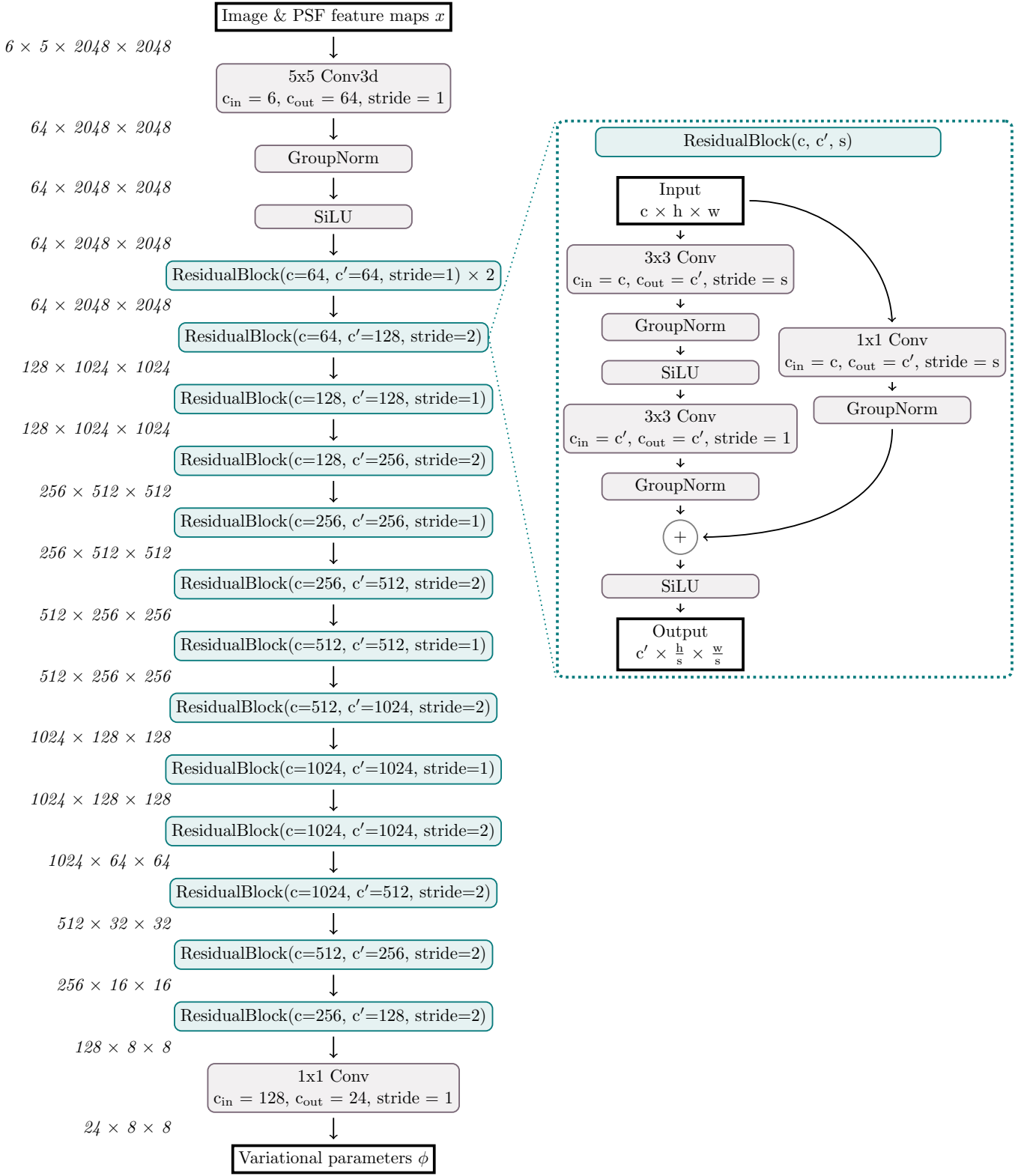
\captionof{figure}{The architecture of our neural network $f_\eta$.}
    \label{fig:architecturediagram}
\end{minipage}


\end{document}